\documentclass[a4paper,12pt]{amsart}
\usepackage{graphicx}  
\usepackage{multicol}  
\makeindex             

\newcommand{\I}{\mathrm i}

\newcommand{\beq}{\begin{equation}
}
\newcommand{\eeq}{\end{equation}}

\newtheorem{theorem}{Theorem}[section]

\theoremstyle{remark}
\newtheorem{remark}[theorem]{Remark}

\numberwithin{equation}{section} \makeatletter
\def\@cite#1#2{#1\if@tempswa , #2\fi}
\def\@biblabel#1{$^{\hbox{\scriptsize{#1}}}$}

\begin{document}

\title[Continuous variable teleportation]{Teleportation of Continuous Quantum Variables: A new Approach}

\author{S. Nagamachi \and E. Br\"uning}

\address[S. Nagamachi]{
Department of Applied Physics and Mathematics, Faculty of
Engineering, The University of Tokushima\\ Tokushima 770-8506,
Japan} \email{shigeaki@pm.tokushima-u.ac.jp}

\address[E. Br\"uning]{School of Mathematical Sciences,
University of KwaZulu-Natal, Private Bag X54001, Durban 4000,
South Africa} \email{bruninge@ukzn.ac.za}

\begin{abstract}
Teleportation of optical field states (as continuous quantum
variables) is usually described in terms of  Wigner functions.  This
is in marked contrast to the theoretical treatment of teleportation
of qubits. In this paper we  show that by using the holomorphic
representation of the canonical commutation relations, teleportation
of continuous quantum variables can be treated in complete analogy
to the case of teleportation of qubits. In order to emphasize this
analogy, short descriptions of the basic experimental schemes both
for teleportation of qubits and of continuous variables are
included. We conclude our paper with a brief discussion of the
effectiveness of our description of continuous variable
teleportation and of the role of localization of quantum states in
teleportation problems.
\end{abstract}

\maketitle

\section{Introduction}
The essence of the quantum information processing consists
in (1) generating quantum entanglements among quantum
systems, (2) controlling the quantum entanglements.
Quantum teleportation contains the essence of
quantum information processing ((1) and (2)).  Quantum
teleportation will play an important role in the realization of  the quantum computer.

 Quantum teleportation is first discovered in 1993 by Bennett et al. [\cite{Be93}]. A few years later two experimental reports on quantum teleportation were published, in 1997 by
Bouwmeester et al. [\cite{Bo97}], and in 1998 by Furusawa et al.
[\cite{Fu98}].

The experiment (by Bouwmeester et al. [\cite{Bo97}]) for the
teleportation of photon states (qubits) is difficult, because the
efficiency of single photon experiments is presently restricted in
principle due to the inability to identify all four Bell states,
and also in practice by the low efficiency of single photon
production and detection. In contrast, the important feature of
the technique used in the experiment for the teleportation of
optical field states by Furusawa et al. [\cite{Fu98}] is its high
efficiency. This is due to the in principle ability to perform the
required joint measurements and the technical maturity of optical
field detection. The experiment of Furusawa et al. [\cite{Fu98}]
is often considered to be the first experimental realization of
quantum teleportation.

The purpose of this lecture is to present teleportation of photon
states and optical field states in a unified way, with emphasis on
a new approach to teleportation of optical field states (quantum
teleportation of continuous quantum variables\index{teleportation
of continuous quantum variables}), based on the holomorhic
representation of the canonical commutation relations instead on
the use of Wigner functions. Here we try to have our lecture self
contained as far as possible (not with respect to the underlying
literature but respect to the arguments which are used).

 The mathematics for the teleportation of photon states (qubits) is the theory of $8 \times 8$ matrices, that is, the theory of 8 dimensional Hilbert spaces. However, in order to describe the
optical field states we need the theory of infinite
dimensional Hilbert spaces. The experiment uses the
entangled states of squeezed laser beams with squeezing
parameter $r$.  Since we can neither generate an infinitely
squeezed $(r = \infty )$ EPR state nor prepare ideal
detectors with efficiency 1, we cannot have complete
teleportation $\psi _{{\rm out}} = \psi _{{\rm in}}$. We
have to measure the quality of the output state $\psi
_{{\rm out}}$. To do so, one often uses the notion of
fidelity\index{fidelity} which is defined by $F(\rho , \sigma ) = {\rm tr \, }\sqrt{\rho ^{1/2}\sigma \rho ^{1/2}}$ for density
operators $\rho $ and $\sigma $.

In the description of teleportation of continuous quantum
variables, one often uses the Wigner function which allows a
compact form for the output state $\psi _{{\rm out}}$ for a given
coherent input state $\psi _{{\rm in}}$ and has an intimate
connection with the fidelity (see [\cite{BK98}]). Such a
formulation sets the theoretical description teleportation of
continuous quantum variables apart from the description of
teleportation of qubits which is based on the use of state vectors
and projective measurements.

In order to understand the essence of the teleportation of
continuous quantum variables and to show the close analogy to
teleportation of quibits, we use the ``holomorphic
representation'' of  the canonical commutation relations (CCR),
\index{canonical commutation relations, CCR} which was introduced
by Bargmann [\cite{Ba61}]. This representation allows to derive
easily explicit formulae for the theoretical description of the
basic operations and objects used in the experiment.

The contents of this lecture can briefly be described as follows:
In Section 2, we recall  quantum teleportation of qubits, in
Section 3, we present briefly the experiment of teleportation of
qubits. In Section 4, we introduce the holomorphic representation
of CCR and determine explicitly the kernels of the mathematical
operations (Bogoliubov transformation) which are needed in the
Section 5 to derive the mathematical realization
 of the devices for the manipulation of photon states (squeezed vacuum state, half-beam splitter, displacement operator).

The theory of laser and parametric oscillator
(amplifier)\index{parametric oscillator} are important for quantum
teleportation but not included in this lecture. We refer for
example to [\cite{SZ97,GZ04}]. In Section 6, we present our
approach to  quantum teleportation of continuous variables. In
Section 7, the experiment made by Furusawa et al. [\cite{Fu98}] is
briefly presented with a discussion of a controversy about this
experiment. Some remarks on the notion of locality\index{locality}
in the theory of quantum teleportation conclude this lecture.

For the convenience of the reader, the first part of an appendix explains the notion of generalized states (which play an essential r\^{o}le in our approach) and mentions some basic properties; the second part gives the detailed proofs for the results presented in Section 4.

\section{Teleportation of qubits}
For the teleportation of qubits
$$\vert \psi \rangle = \alpha \vert 0\rangle + \beta \vert 1\rangle
\Leftrightarrow
 \left( \begin{array}{c} \alpha \\{} \beta \end{array} \right) \in  \mathbb C^{2}, \  \vert 0\rangle
 = \left( \begin{array}{c} 1\\{} 0\end{array} \right) , \vert 1\rangle
 = \left( \begin{array}{c} 0\\{} 1\end{array} \right) $$
 the following single qubit gates are used:
$$X = \left(\begin{array}{ccc} 0&1\\{} 1&0\end{array} \right),
\  X \left( \begin{array}{c} \alpha \\{} \beta \end{array} \right)=
\left(\begin{array}{c} \beta \\{} \alpha \end{array} \right), \
X(\vert 0\rangle, \vert 1\rangle) = (\vert 1\rangle, \vert 0\rangle),$$
$$ Y = \left( \begin{array}{ccccccc} 0&-\I\\{} \I&0\end{array} \right), \
 Z = \left( \begin{array}{ccccc} 1&0\\{} 0&-1\end{array} \right), \
 H=\frac{1}{\sqrt{2}}\left(\begin{array}{ccc} 1&1\\{} 1&-1\end{array}\right)=(X+Z)/\sqrt{2}, $$
$$ H(\vert 0\rangle, \vert 1\rangle)=\frac{1}{\sqrt{2}}(\vert 0\rangle +
\vert 1\rangle, \vert 0\rangle - \vert 1\rangle).$$
The matrix $H$ is called the Hadamard gate{\sf
\index{Hadamard gate}}.\newline \indent In order to treat
multi-qubits, we must consider  composite systems.
\newline {\bf Axiom of composite system}: The state space
of the composite physical system is the tensor product of
the state spaces of the component physical systems.
Moreover, if we have systems numbered 1 through $n$, and
system number $i$ is prepared in the state $\vert \psi
_{i}\rangle $, then the state of the total system is $\vert
\psi _{1}\rangle \otimes \vert \psi _{2}\rangle  \otimes
\cdots \otimes \vert \psi _{n}\rangle $. \vskip 12pt
\noindent
The two qubit gate (Controlled-Not gate)\index{controlled-not gate} $M_{{\rm CNOT \, }}$ will play an important role in our lecture. This gate acts on the basis vectors
$$ \vert ij\rangle = \vert i\rangle \otimes \vert j\rangle =
\vert i\rangle \vert j\rangle$$
as follows:
$$ M_{{\rm CNOT \,}}(\vert 00\rangle,
\vert 01\rangle,\vert 10\rangle, \vert 11\rangle) = (\vert 00\rangle,
\vert 01\rangle,\vert 11\rangle, \vert 10\rangle).$$

\subsection{ EPR pair, Bell states}
The essence of quantum information science is quantum
entanglement and its manipulation.  The entangled states
$\vert \beta _{ij}\rangle $ called EPR pair\index{EPR pair}
or Bell states\index{Bell states} are created by using
Hadamard gate $H$ and controlled-not gate $M_{\rm CNOT}$:
\begin{equation}\label{Bell-basis}
\begin{aligned}
M_{{\rm CNOT\, }}(H\otimes I)\vert 00\rangle&=\frac{1}{\sqrt{2}}
M_{{\rm CNOT \, }}(\vert 0\rangle + \vert 1\rangle)\otimes \vert
0\rangle=\frac{1}{\sqrt{2}}(\vert 00\rangle +\vert 11\rangle)=
\vert \beta_{00}\rangle,\\
M_{{\rm CNOT\,}}(H\otimes I)\vert 01\rangle&=\frac{1}{\sqrt{2}}
M_{{\rm CNOT \,}}(\vert 0\rangle + \vert 1\rangle)\otimes \vert 1\rangle
= \frac{1}{\sqrt{2}}(\vert 01\rangle +\vert 10\rangle)=\vert \beta_{01}
\rangle,\\
M_{{\rm CNOT \,}}(H\otimes I)\vert 10\rangle&=\frac{1}{\sqrt{2}}
M_{{\rm CNOT \,}}(\vert 0\rangle - \vert 1\rangle)\otimes \vert 0\rangle
= \frac{1}{\sqrt{2}}(\vert 00\rangle -\vert 11\rangle) = \vert \beta_{10}
\rangle,\\
M_{{\rm CNOT \,}}(H\otimes I)\vert 11\rangle&=\frac{1}{\sqrt{2}}
M_{{\rm CNOT \,}}(\vert 0\rangle - \vert 1\rangle)\otimes \vert 1\rangle
=\frac{1}{\sqrt{2}}(\vert 01\rangle -\vert 10\rangle)=\vert\beta_{11}\rangle .\\
\end{aligned}
\end{equation}
\begin{remark} \label{rem2.1}
The canonical basis $\vert ij\rangle $ can be written in terms of the Bell
basis\index{Bell basis} $\vert \beta _{ij}\rangle $ as
follows:
\begin{equation} \label{inverse-Bell}
\begin{aligned}
 &\vert 00\rangle = \frac{1}{\sqrt{2}}(\vert \beta _{00}\rangle  + \vert \beta _{10}\rangle ),\quad \  \vert 11\rangle  = \frac{1}{\sqrt{2}}(\vert \beta _{00}\rangle  - \vert \beta _{10}\rangle ),\\
 &\vert 01\rangle  = \frac{1}{\sqrt{2}}(\vert \beta _{01}\rangle  + \vert \beta _{11}\rangle ),\quad \  \vert 10\rangle  = \frac{1}{\sqrt{2}}(\vert \beta _{01}\rangle  - \vert \beta _{11}\rangle ).
\end{aligned} \end{equation}
\end{remark}
\subsection{ Description of quantum measurement}
{\bf Axiom of quantum measurement}:
Quantum measurements are described by a collection $\{
M_{m}\} $ of projection operators which appear in the
spectral decomposition of the observable
$$       M = \sum _{m} m M_{m}.$$
The index $m$ refers to the measurement outcomes that may
occur in the experiment.  If the state of the quantum
system is $\vert \psi \rangle $ ($\Vert \vert \psi \rangle
\Vert  = 1$) immediately before the measurement then the
probability that result $m$ occurs is given by
$$       p(m) = \langle \psi \vert M_{m}\vert \psi \rangle ,$$
and the state of the system after such an ideal measurement is
$$       \frac{M_{m}\vert \psi \rangle }{\sqrt{\langle \psi \vert M_{m}\vert \psi \rangle }}.$$
The projection operators satisfy the completeness equation,
$$       \sum _{m} M_{m} = I = \textrm{Identity operator on the state space}.$$
\subsection{Quantum teleportation}
Now we can describe the process of quantum teleportation,
which is illustrated in Fig.~1.
\begin{figure}[!h]
  \begin{center}
\includegraphics[width=12cm]{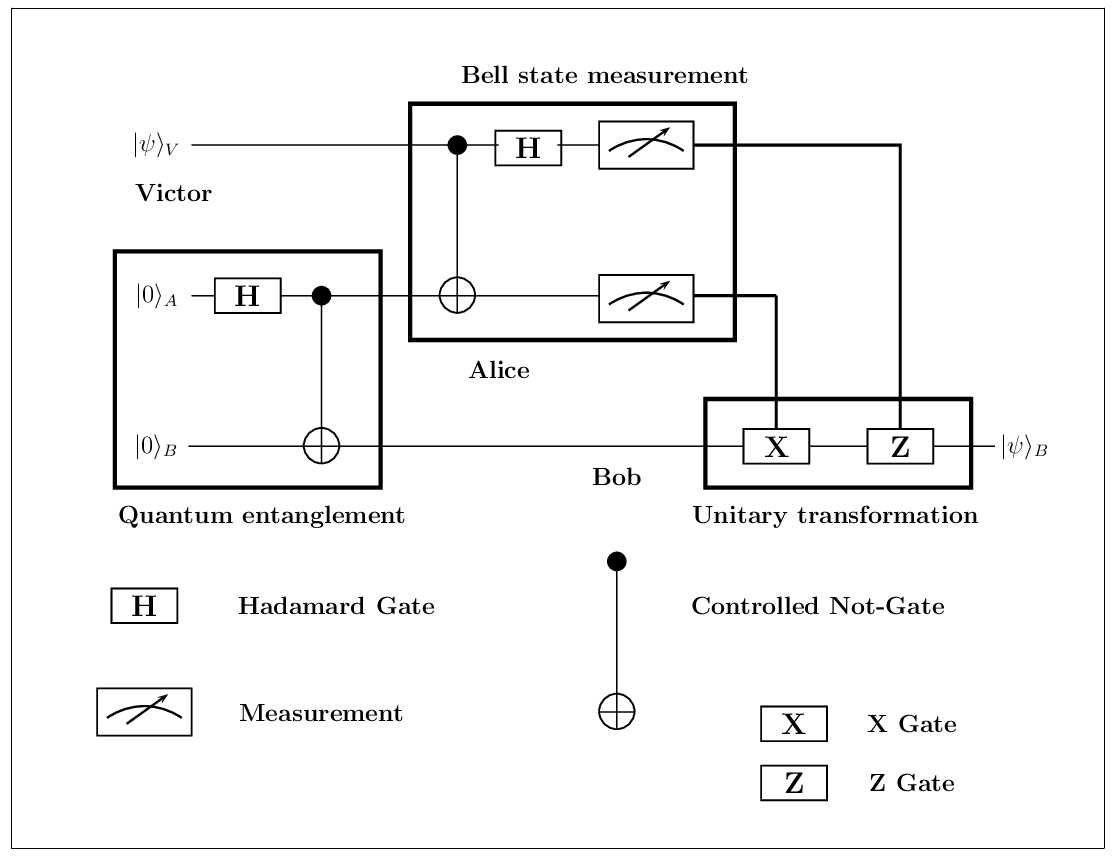}
    \caption{Quantum circuit of
teleportation. Victor's state $|\psi\rangle_V$ is given to
Alice and reproduced at Bob's laboratory.}
    \label{fig:teleportqu3}
  \end{center}
\end{figure}

\begin{enumerate}
\item[(1)] Alice and Bob prepare an EPR pair $\vert \beta _{00}\rangle _{AB}$
$$       \vert \beta _{00}\rangle _{AB}
 = \frac{1}{\sqrt{2}}(\vert 0\rangle _{A} \otimes \vert 0\rangle _{B} + \vert 1\rangle _{A} \otimes \vert 1\rangle _{B})
 = \frac{1}{\sqrt{2}}(\vert 0\rangle _{A} \vert 0\rangle _{B} + \vert 1\rangle _{A} \vert 1\rangle _{B})$$
by applying a Hadamard gate and a controlled-not gate to the
state $\vert 0\rangle _{A} \otimes \vert 0\rangle _{B}$,
the composite state of Alice's state $\vert 0\rangle _{A}$
and Bob's state $\vert 0\rangle _{B}$.  Now Alice and Bob
share the state $\vert  \beta _{00}\rangle _{AB}$ ($\vert 0
\rangle _{A}$, $\vert 1\rangle _{A}$ are Alice's states,
and $\vert 0\rangle _{B}$, $\vert  1\rangle _{B}$ are Bob's
states).\newline Victor gives Alice a state $\vert \psi
\rangle _{V}$ ($\Vert \vert \psi \rangle _{V}\Vert  = 1$)
to send to Bob. The state of the total system is
$$       \vert \psi _{0}\rangle  = \vert \psi \rangle _{V} \otimes \vert \beta _{00}\rangle _{AB} =
(\alpha \vert 0\rangle _{V} + \beta \vert 1\rangle
_{V})\otimes \frac{1}{\sqrt{2}}(\vert 0\rangle _{A}\vert
0\rangle _{B} + \vert 1\rangle _{A}\vert 1\rangle _{B}).$$
The state $\vert \psi _{0}\rangle  = \vert \psi \rangle
_{V}\otimes \vert \beta _{00}\rangle _{AB}$ can be rewitten as:
$$ \vert \psi _{0}\rangle  = (\alpha \vert 0\rangle _{V} + \beta \vert 1\rangle _{V})\otimes \frac{1}{\sqrt{2}}(\vert 0\rangle _{A}\vert 0\rangle _{B} + \vert 1\rangle _{A}\vert 1\rangle _{B})$$
$$  = \frac{1}{\sqrt{2}}[\alpha \vert 0\rangle _{V}\vert 0\rangle _{A}\vert 0\rangle _{B} + \alpha \vert 0\rangle _{V}\vert 1\rangle _{A}\vert 1\rangle _{B}$$
$$ + \beta \vert 1\rangle _{V}\vert 0\rangle _{A}\vert 0\rangle _{B} + \beta \vert 1\rangle _{V}\vert 1\rangle _{A}\vert 1\rangle _{B}]$$
$$= \frac{1}{2} [\alpha (\vert \beta _{00}\rangle _{VA} + \vert \beta _{10}\rangle _{VA})\vert 0\rangle _{B} + \alpha  (\vert \beta _{01}\rangle _{VA} + \vert \beta _{11}\rangle _{VA})\vert 1\rangle _{B}$$
$$+ \beta (\vert \beta _{01}\rangle _{VA} - \vert \beta _{11}\rangle _{VA})\vert 0\rangle _{B} + \beta (\vert \beta _{00}\rangle _{VA} - \vert \beta _{10}\rangle _{VA})\vert 1\rangle _{B}]$$
$$    = \frac{1}{2} [\vert \beta_{00}\rangle_{VA}(\alpha \vert 0\rangle_{B} + \beta
\vert 1\rangle_{B}) + \vert \beta_{01}\rangle_{VA}(\alpha \vert 1\rangle_{B} +
\beta \vert 0\rangle_{B})$$
\begin{equation} \label{2.1}
 + \vert \beta_{10}\rangle_{VA}(\alpha \vert 0\rangle_{B} - \beta \vert 1\rangle_{B})
 + \vert \beta_{11}\rangle_{VA}(\alpha \vert 1\rangle_{B} - \beta \vert 0\rangle_{B})].
 \end{equation}
\item[(2)] Alice performs the Bell-state measurement\index{Bell-state measurement}, a measurement
which determines $\vert \beta _{ij}\rangle $, and the state
of the system after the measurement is
$$    \left\{ \begin{array}{cc}
     \vert \beta _{00}\rangle _{VA}(\alpha \vert 0\rangle _{B}+\beta \vert 1\rangle _{B}) & {\rm if \, }(ij) = (00) \\
     \vert \beta _{01}\rangle _{VA}(\alpha \vert 1\rangle _{B}+\beta \vert 0\rangle _{B}) & {\rm if \, }(ij) = (01) \\
     \vert \beta _{10}\rangle _{VA}(\alpha \vert 0\rangle _{B}-\beta \vert 1\rangle _{B}) & {\rm if \, }(ij) = (10) \\
     \vert \beta _{11}\rangle _{VA}(\alpha \vert 1\rangle _{B}-\beta \vert 0\rangle _{B}) & {\rm if \, }(ij) = (11)
     \end{array} \right.  .$$
\item[(3)] Alice  sends the classical information $(ij)$ to
Bob. Then Bob sends Bob's qubit through $I, X, Z, XZ$
according to the result $(00), (01), (10), (11)$, obtaining
$\alpha \vert 0\rangle _{B} + \beta \vert 1\rangle
_{B}$.\newline
\item[(4)] Since the operator $(H\otimes I)M_{{\rm CNOT \,}}$ is the inverse of
the operator $M_{{\rm CNOT\,}}(H\otimes I)$, $(H\otimes I)M_{{\rm CNOT\,}}$
 sends the Bell basis $\vert \beta_{ij}\rangle_{VA}$ to the
canonical basis $\vert ij\rangle_{VA}=\vert i\rangle_{V}\otimes \vert
j\rangle_{A}$ of product states.
Thus Alice's Bell-state measurement is performed by sending
the Bell basis to the canonical basis $\vert ij\rangle
_{VA}$ and making the measurement determining $\vert
ij\rangle _{VA}$.  In practice, Alice sends her qubits
through a CNOT gate, and then the first qubit through a
Hadamard gate $H$, obtaining
$$ \frac{1}{2}[\vert 00\rangle _{VA}(\alpha \vert 0\rangle _{B} + \beta \vert 1\rangle _{B}) + \vert 01\rangle _{VA}(\alpha \vert 1\rangle _{B}+\beta \vert 0\rangle _{B})$$
\begin{equation} \label{2.2}
  + \vert 10\rangle _{VA}(\alpha \vert 0\rangle _{B} - \beta
\vert 1\rangle _{B}) + \vert 11\rangle _{VA}(\alpha \vert
1\rangle _{B} - \beta \vert 0\rangle _{B})].
\end{equation}
Let $M_{ij} = \vert ij\rangle _{VA}{}_{VA}\langle ij\vert \otimes
I_{B}$, $(i, j = 1, 2)$.  Then these projection operators satisfy the completeness equation.\newline
\item[(5)] Alice performs a measurement
$\{ M_{ij}\} $, i.e., measures the observable $M = \sum _{i,j =
0}^{1}$ $(2i+j)M_{ij}$.  The probability that result $(ij)$ occurs
(equivalently $m = 2i+j$) is
$$ p(ij) = \langle \psi _{2}\vert M_{ij}\vert \psi _{2}\rangle  = \frac{1}{4},$$
and the state of the system after the measurement is
$$ \frac{M_{ij}\vert \psi _{2}\rangle }{\sqrt{\langle \psi _{2}\vert M_{ij}\vert \psi _{2}\rangle }}
  = \left\{ \begin{array}{ccc}
     \vert 00\rangle _{VA}(\alpha \vert 0\rangle _{B}+\beta \vert 1\rangle _{B}) & {\rm if \, }(ij) = (00) & m = 0 \\
     \vert 01\rangle _{VA}(\alpha \vert 1\rangle _{B}+\beta \vert 0\rangle _{B}) & {\rm if \, }(ij) = (01) & m = 1 \\
     \vert 10\rangle _{VA}(\alpha \vert 0\rangle _{B}-\beta \vert 1\rangle _{B}) & {\rm if \, }(ij) = (10) & m = 2 \\
     \vert 11\rangle _{VA}(\alpha \vert 1\rangle _{B}-\beta \vert 0\rangle _{B}) & {\rm if \, }(ij) = (11) & m = 3
     \end{array} \right.  .$$
     \end{enumerate}

\nopagebreak
\section{Experiment of the qubit teleportation}
In this section we discuss the experiment performed by Bouwmeester
et al. in 1997 [\cite{Bo97}] using qubit states. Two pairs of
entangled photons  are generated by a polarized non-degenerate
parametric process.  Let $\vert 0 \rangle $ represent the
horizontally polarized single photon state
$|\leftrightarrow\rangle $ and $\vert 1 \rangle $ the vertically
polarized single photon state $|\updownarrow \rangle $.

For these polarization states we calculate the Bell basis according to (\ref{Bell-basis}).
Then the state $\vert \psi
_{0}\rangle  = \vert \psi \rangle _{V} \otimes \vert \beta
_{00}\rangle _{AB}$ is rewritten as follows, (see (\ref{2.1})):
$$ |\psi_0\rangle = \frac{1}{2} [\vert \beta _{00}\rangle _{VA}(\alpha \vert 0\rangle _{B} + \beta \vert 1\rangle _{B})
  + \vert \beta _{01}\rangle _{VA}(\alpha \vert 1\rangle _{B} + \beta \vert 0 \rangle _{B})$$
$$  + \vert \beta _{10}\rangle _{VA}(\alpha \vert 0\rangle _{B} - \beta \vert 1\rangle _{B}
  + \vert \beta _{11}\rangle _{VA}(\alpha \vert 1\rangle _{B} - \beta \vert 0\rangle _{B})].$$
A Bell-state measurement is possible for the state $\vert
\beta _{11}\rangle _{VA}$, under the condition that there
are at most one photon on each mode of Victor and Alice.
Let $a_{Vj}, a_{Ak}$ be annihilation operators of
polarization $j, k = \leftrightarrow , \updownarrow $.  The
half-beam splitter causes the Bogoliubov transformation (see
(\ref{dev4}))
$$    b_{0j} = \frac{1}{\sqrt{2}} (a_{Vj} + a_{Aj}), \ b_{1j} = \frac{1}{\sqrt{2}} (-a_{Vj} +
a_{Aj}). $$  Let $\vert \Omega \rangle $ be the vacuum for
$a_{Vj}, a_{Ak}$. Then we have
$$   \langle \Omega \vert b_{1k} b_{0j} = \frac{1}{2}  \langle \Omega \vert (-a_{Vk} + a_{Ak}) (a_{Vj} + a_{Aj}). $$
We assume that there is at most one photon in each mode
(Victor's mode or Alice's mode). On such a condition, we
can perform the Bell-state measurement by the simultaneous
photon counting after the half-beam splitter.  For such a
state $\vert \psi \rangle $, we can ignore $a_{Vk}a_{Vj}$
and $a_{Ak}a_{Aj}$ we have
$$     \langle \Omega \vert b_{1k} b_{0j} \vert \psi \rangle = \langle \Omega \vert (-a_{Aj}a_{Vk} + a_{Ak}a_{Vj}) \vert \psi \rangle =
\pm \frac{1}{\sqrt{2}} {}_{VA}\langle \beta _{11} \vert
\psi \rangle $$
 for $j \neq k$, and $\langle \Omega \vert b_{1k} b_{0j} \vert \psi \rangle = 0$ for $j = k$, where we used
the fact
$$ (-a_{Vk}^{\dagger} a_{Aj}^{\dagger} + a_{Vj}^{\dagger} a_{Ak}^{\dagger})\vert \Omega \rangle = \pm \frac{1}{\sqrt{2}} \vert \beta _{11}\rangle _{VA}. $$
This shows that the simultaneous photon detection is
equivalent to ${}_{VA}\langle \beta _{11} \vert $, that is,
equivalent to the Bell-state measurement. However, this
measurement cannot identify the other three Bell states, and has
a fatal drawback.  The photon counting technique of today
cannot distinguish whether only one photon is coming or
more than two photons are coming simultaneously. So, if two
photons come in Victor's mode simultaneously, we cannot
neglect the term $a_{Vk}a_{Vj}$, the measurement is not the
Bell-state measurement.

\section{Holomorhic Representation of CCR}
The ``holomorphic representation'' of  the canonical commutation relations (CCR)
\index{canonical commutation relations, CCR}
 was introduced by Bargmann [\cite{Ba61}] for the finite dimensional case. In the
case of infinitely many degrees of freedom it was introduced by
Segal [\cite{Se62}] (see also [\cite{Se78}]). One of the most
famous applications of this representation we find in the book
[\cite{Be66}] by Berezin,  where it has the interesting
counterpart to the Fermion case, i.e., canonical anti-commutation
relations. The canonical anti-commutation relation\index{canonical
anti-commutation relation} has a representation which is similar
to the holomorphic representation where however the field of
complex numbers is replaced by the Grassmann algebra. Such a
representation is now popular under the name of Berezin calculus.
Another important application of the holomorhic representation is
given in the book of Faddeev and Slanov [\cite{FS80}].

\subsection{The holomorphic (Bargmann) representation of the CCR}
\noindent We develop the representation of canonical
commutation relation (CCR) called holomorphic
representation, which seems to be quite useful for quantum
optics and quantum teleportation of continuous variables.
\vskip 12pt \noindent The operator of multiplication $q$
and the differentiation $p = -\I d/dq$ in $L^{2}(\mathbb R) =
L^{2}(\mathbb R, dq)$ satisfies the commutation relation:
$$       [q, p] = qp - pq =  \I I$$
on a suitable subspace of $L^{2}(\mathbb R)$.  Introduce
the operators
$$       a = (q + d/dq)/\sqrt{2} = (q + \I p)/\sqrt{2}, \  a^{\dagger} = (q - \I p)/\sqrt{2}$$
then
$$       [a, a^{\dagger}] = 1, \  q = (a + a^{\dagger})/\sqrt{2}, \  p = (a - a^{\dagger})/\sqrt{2}\I.$$
This representation of the commutation relation is called
the {\it Schr\"odinger representation}\index{Schr\"odinger representation}.  The function $f(q)
= e^{-q^{2}/2}$ is a solution of the equation
$$      0 = \sqrt{2}a f(q) = (q + d/dq)f(q).$$
We consider the space $L^{2}(\mathbb R, e^{-q^{2}}
dq/\sqrt{\pi })$ and the unitary operator $U$
$$      U: L^{2}(\mathbb R, e^{-q^{2}} dq/\sqrt{2\pi }) \ni  g(q) \rightarrow  (\pi )^{-1/2}g(q) e^{-q^{2}/2} \in  L^{2}(\mathbb R, dq).$$
Then we have
$$      U^{\dagger}qU = q, \  U^{\dagger}(d/dq)U = d/dq - q,$$
$$      b = U^{\dagger}aU = U^{\dagger}(q + d/dq)U/\sqrt{2} = 2^{-1/2}d/dq,$$
$$      b^{\dagger} = U^{\dagger}a^{\dagger}U = U^{\dagger}(q - d/dq)U/\sqrt{2} = 2^{-1/2}(2q - d/dq).$$
This representation $L^{2}(\mathbb R, e^{-q^{2}}
dq/\sqrt{2\pi })$, $b = 2^{-1/2}d/dq$, $b^{\dagger} = 2^{-1/2}(2q
- d/dq)$ is called the {\it modified Schr\"odinger
representation}\index{modified Schr\"odinger representation}.\newline Next we consider
$$  u = x + \I y, \  L^{2}(\mathbb C, d\mu ), \  d\mu  = e^{-\bar{u}  u} \frac{d\bar{u}  du}{2\pi \I} =e^{-(x^{2}+y^{2})}\frac{dxdy}{\pi }, \  \int _{\mathbb C} d\mu = 1,$$
and its subspace ${\mathcal H} $ generated by holomorphic
functions of $\bar{u}  $ (anti-holomorphic functions).  For
$f, g \in  {\mathcal H} $ define the inner product by
$$ \langle f\vert g\rangle  = \int \overline{f(\bar{u}  )}g(\bar{u}  )e^{-\bar{u}  u} \frac{d\bar{u}  du}{2\pi \I}.$$
Then  the multiplication operator $f(\bar{u}  ) \rightarrow
\bar{u}   f(\bar{u}  )$ and the differential operator
$f(\bar{u}  ) \rightarrow  \partial /\partial \bar{u}
f(\bar{u}  )$ are adjoint to each other:
$$  \int \overline{f(\bar{u})} \{ \partial /\partial \bar{u}  g(\bar{u} )\} e^{-\bar{u} u} \frac{d\bar{u}  du}{2\pi \I}
= -\int \partial /\partial \bar{u}  \{ e^{-\bar{u}  u} \overline{f(\bar{u}  )}\} g(\bar{u}  ) \frac{d\bar{u}  du}{2\pi \I}$$
$$ = \int  \overline{f(\bar{u}  )} ug(\bar{u}  )e^{-\bar{u}  u} \frac{d\bar{u}  du}{2\pi \I} = \int  \overline{\bar{u}   f(\bar{u}  )} g(\bar{u}  )e^{-\bar{u}  u} \frac{d\bar{u}  du}{2\pi \I},$$
where we used the relation $\partial /\partial \bar{u}
\overline{f(\bar{u}  )} = 0$ because $\overline{f(\bar{u})}$ is a holomorphic function of $u$.  If we put
$$ a = \partial /\partial \bar{u} =(1/2)(\partial /\partial x + \I \partial /\partial y), \  a^{\dagger} = \bar{u}  ,$$
then $a$ and $a^{\dagger}$ satisfy the commutation relation $[a,a^{\dagger}] = 1$. We call this representation the {\it holomorphic representation\index{holomorphic
representation}}.  In this representation, $\{ \bar{u}
^{n}/\sqrt{n!}\} _{n=0}^{\infty }$ is an orthonormal basis.
In fact, let $m \leq  n$.  Then we have
$$ \int \frac{u^{m}}{\sqrt{m!}} \frac{\bar{u}^{n}}{\sqrt{n!}} e^{-\bar{u}u} \frac{d\bar{u}  du}{2\pi \I}
= \int \frac{u^{m}}{\sqrt{m!}} \left(\frac{(-1)^{n}\partial ^{n}}{\sqrt{n!} \partial u^{n}}\right)  e^{-\bar{u} u} \frac{d\bar{u} du}{2\pi \I}$$
$$= \int \left(\frac{\partial ^{n}}{\sqrt{n!} \partial u^{n}} \frac{u^{m}}{\sqrt{m!}}\right) e^{-\bar{u}u} \frac{d\bar{u}  du}{2\pi \I} = \delta_{mn},$$
and if $m > n$,
$$ \int \frac{u^{m}}{\sqrt{m!}} \frac{\bar{u}^{n}}{\sqrt{n!}} e^{-\bar{u} u} \frac{d\bar{u}  du}{2\pi \I}
= \int \left( \frac{\partial^{m}}{\sqrt{m!} \partial \bar{u}^{m}} \frac{\bar{u}^{n}}{\sqrt{n!}}\right)  e^{-\bar{u}u} \frac{d\bar{u}du}{2\pi \I} = 0.$$
Therefore, if $f(\bar{u}) = \sum_{n=0}^{\infty } a_{n} \bar{u} ^{n}$ then $\Vert f\Vert ^{2} = \sum _{n=0}^{\infty } \vert a_{n}\vert ^{2} n!$. The coherent
state\index{coherent state} is an eigen state of the annihilation operator. Also in the holomorphic
representation,
$$ \frac{\partial}{\partial \bar{u}} f(\bar{u}) = \alpha f(\bar{u}), \  f(\bar{u}) = C e^{\alpha \bar{u}} = C \sum_{n=0}^{\infty} \frac{(\alpha \bar{u})^{n} }{n!},$$
$$ C = e^{-\vert \alpha \vert ^{2}/2} \Rightarrow  \Vert f\Vert^{2}= \vert C\vert^{2} \sum _{n=0}^{\infty } \frac{\vert \alpha \vert ^{2n}}{n!} = \vert C\vert^{2} e^{\vert \alpha \vert^{2}} = 1.$$
Hence, $f$ is normalized by choosing $C = e^{-\vert \alpha \vert ^{2}/2}$.  Furthermore, in the holomorphic
representation, the integral kernel $K(\bar{u} , v)$ of
the identity operator is
$$ K(\bar{u}, v) = e^{\bar{u} v}.$$
In fact, for $\alpha  \in  \mathbb C$,
$$ \int e^{\alpha v} \bar{v}^{n} e^{-\bar{v}v} \frac{d\bar{v} dv}{2\pi \I} = \int \sum _{k=0}^{\infty }\frac{(\alpha v)^{k}}{k!} \bar{v}^{n} e^{-\bar{v} v} \frac{d\bar{v}  dv}{2\pi \I} = \alpha^{n}$$
implies
\begin{equation} \label{4.1}
  \int e^{\alpha v} f(\bar{v} ) e^{-\bar{v} v} \frac{d\bar{v}
dv}{2\pi \I} = \int e^{\alpha v} \sum_{n=0}^{\infty }
a_{n}\bar{v}^{n} e^{-\bar{v} v} \frac{d\bar{v}  dv}{2\pi \I} = \sum_{n=0}^{\infty } a_{n}\alpha^{n} = f(\alpha ).
\end{equation}
Because of symmetry in $v$ and $\bar{v}  $, one also has
\begin{equation} \label{4.2}
     \int e^{\alpha \bar{v} } f(v) e^{-\bar{v}v} \frac{d\bar{v}  dv}{2\pi \I} = f(\alpha ).
\end{equation}
The equality
$$ a^{\dagger m}a^{n} f(\bar{u}) = \bar{u}^{m} \frac{\partial ^{n}}{\partial \bar{u}^{n}} \int e^{\bar{u} v} f(\bar{v}) e^{-\bar{v} v} \frac{d\bar{v} dv}{2\pi \I} = \int \bar{u}^{m}v^{n}e^{\bar{u} v} f(\bar{v} ) e^{-\bar{v} v} \frac{d\bar{v} dv}{2\pi \I} $$
shows that the integral kernel of the normal ordered
monomial $a^{\dagger m}a^{n}$ of $a^{\dagger}$ and $a$ is $\bar{u}^{m}v^{n}e^{\bar{u} v}$.\\
The holomorphic representation for the case of {\it  $n$ degrees of freedom} is briefly described below:\\
The variables, the Hilbert space and the measure now are:
$$\bar{u}= (\bar{u}_{1}, \ldots, \bar{u}_{n}), \  L^{2}(\mathbb C^{n}, d\mu _{n}), \  d\mu _{n} = \prod _{j=1}^{n} e^{-\bar{u}_{j}u_{j}} \frac{d\bar{u}_{j}du_{j}}{2\pi \I}.$$
\subsection{Integral kernels of basic operations in the holomorhic representation}
For the mathematical description of the devices for the manipulation of photon states we need
several results for the kernels of various linear transformations of the basic creation and annihilation operators. The relevant results are stated in this subsection. The proofs are contained in the appendix.

These results are special cases of results given in [\cite{Be66}]
where the infinite dimensional counter part is given. The proofs
of these results in [\cite{Be66}] are based on the use of
functional integration. Though the basic strategy is the same as
in [\cite{Be66}], the proofs given in our appendix only uses
elementary mathematical tools so that these result become more
easily accessible.
\begin{theorem} \label{thm4.1}
The linear canonical transformation\index{linear canonical
transformation}
\begin{equation} \label{inhomo}
       b_{j} = a_{j} + f_{j}, \  b_{j}^{\dagger} = a_{j}^{\dagger} + \bar{f}_{j}
\end{equation}
is implemented by the unitary operator $U$ whose integral
kernel $U(\bar{u},v)$ is
$$  U(\bar{u},v) = c \exp \sum _{j=1}^{n} (\bar{u}_{j}v_{j}
+ v_{j}\bar{f}_{j} - \bar{u}_{j}f_{j}), \  c = \theta
\exp \left\{(-1/2)\sum _{j=1}^{n} \bar{f}_{j}f_{j}\right\} , \ \vert \theta \vert  = 1.
$$
\end{theorem}
\vskip 12pt
\begin{theorem} \label{thm4.2}
The linear canonical transformation
$$ b_{j} = \sum_{k=1}^{n}(\Phi_{jk}a_{k} + \Psi_{jk}a^{\dagger}_{k}), \  b_{j}^{\dagger} = \sum_{k=1}^{n}(\bar{\Phi }  _{jk}a_{k}^{\dagger} + \bar{\Psi}_{jk}a_{k})$$
is implemented by the unitary operator $U$ whose integral
kernel $U(\bar{u},v)$ is given by
$$ U(\bar{u},v) = c \exp \left( \frac{1}{2} (v \  \bar{u} ) \left( \begin{array}{ccc} A^{11}&A^{12}\\{} A^{21}&A^{22}\end{array} \right) \left( \begin{array}{c} v\\{} \bar{u} \end{array} \right) \right) , \  A^{jk} = {}^tA^{kj},$$
$$ A^{22} = -\Phi ^{-1}\Psi , \  A^{21} = \Phi ^{-1}, \  A^{11} = \bar{\Psi }  \Phi ^{-1},
 c = \theta  (\det \Phi \Phi ^{\dagger})^{-1/4}, \  \vert \theta
\vert  = 1.$$
\end{theorem}
\vskip 12pt \noindent
\begin{theorem} \label{thm4.3}
Let $f_{j}$ be complex numbers and
$$ H = \sum _{j=1}^{n} (f_{j}a_{j}^{\dagger} + \bar{f}_{j}a_{j})$$
be a self-adjoint operator.  Then the kernel function
$U(\bar{u}, v)$ of $e^{itH}$ is given by
\begin{equation} \label{4.4}
     U(\bar{u}, v) = c \exp \sum_{j=1}^{n} (\bar{u}_{j}v_{j}
     + it\bar{f}_{j} v_{j} + itf_{j}\bar{u}_{j}), \ c = \exp
\left\{ - \frac{1}{2} t^{2} \sum_{j=1}^{n} \bar{f}
_{j}f_{j}\right\}.
\end{equation}
\end{theorem}

\begin{theorem} \label{thm4.4}
Let $B$ and $C$ be $n \times n$ matrices, and
$$ H = \frac{1}{2} (a^{\dagger}Ba^{\dagger} + a\bar{B}  a + 2a^{\dagger}Ca)$$
be a self-adjoint operator.  Further, denote
$$ {\mathcal A}  = \left( \begin{array}{ccccc} -C&-B\\{} \bar{B}  &\bar{C}  \end{array} \right) , \  e^{it{\mathcal A} } = \left( \begin{array}{ccc} \Phi &\Psi \\{} \bar{\Psi }  &\bar{\Phi }  \end{array} \right) .$$
Then the kernel function $U(\bar{u},v)$ of $e^{itH}$ is
given by
\begin{equation} \label{4.7}
    U(\bar{u}, v) = c \exp \left( \frac{1}{2} (v \  \bar{u}  )
\left( \begin{array}{ccc} A^{11}&A^{12}\\{} A^{21}&A^{22}\end{array} \right)
  \left( \begin{array}{c} v\\{} \bar{u}  \end{array} \right) \right) , \  A^{jk} = {}^tA^{kj}
\end{equation}
with
\begin{equation} \label{4.8}
   A^{22} = -\Phi ^{-1}\Psi , \  A^{21} = \Phi ^{-1}, \  A^{11} =
\bar{\Psi }  \Phi ^{-1}, \ c = (\det \Phi e^{itC})^{-1/2}.
\end{equation}
\end{theorem}

\section{Photon states and devices for manipulating photon states}
We consider a self-adjoint operator
$$ H_{{\rm laser \,}} = \I(\bar{\alpha} a - \alpha a^{\dagger})$$
and an operator $D(\alpha) = e^{\I H_{{\rm laser \,}}(\alpha)}$ called
the displacement operator\index{displacement operator} which
generates from the vacuum state a state called the {\it coherent
state} which is considered to represent the laser-beam state. In
order to have the kernel $U(\bar{u}  , v)$ of the operator
$e^{\I H_{{\rm laser \, }}(g)}$, we use Theorem \ref{thm4.3}, i.e., $n
= 1$ and $f = -\I\alpha $. The kernel $U(\bar{u},v)$ of the
operator $D(\alpha )$ therefore is
$$ U(\bar{u},v)=e^{-\vert \alpha \vert^{2}/2} \exp\{\bar{u} v -
\bar{\alpha } v + \alpha \bar{u} \} .$$
In the holomorphic representation, its action on a state
can be calculated explicitly
$$(D(\alpha)f)(\bar{u})= \int U(\bar{u}, v) f(\bar{v}) e^{-\bar{v} v}
 \frac{d\bar{v}  dv}{2\pi \I} $$
$$= e^{-\vert \alpha \vert^{2}/2} \int \exp \{\bar{u}  v - \bar{\alpha}
 v+\alpha \bar{u}\}f(\bar{v})e^{-\bar{v} v} \frac{d\bar{v}dv}{2\pi \I}$$
$$= e^{-\vert \alpha \vert^{2}/2} e^{\alpha \bar{u}} \int \exp \{(\bar{u}
- \bar{\alpha} )v\} f(\bar{v}) e^{-\bar{v}v} \frac{d\bar{v}dv}{2\pi \I} $$
\begin{equation} \label{dev1}
 = e^{-\vert \alpha \vert ^{2}/2} e^{\alpha \bar{u}} f(\bar{u}
- \bar{\alpha}).
\end{equation}
If $f(\bar{u}) = 1$ which corresponds to the vacuum $\vert
0\rangle $, then
$$\vert \alpha \rangle = D(\alpha)\vert 0\rangle = e^{-\vert \alpha
\vert /2} e^{\alpha \bar{u}}$$
is the coherent state\index{coherent state} and satisfies
the following completeness relation\index{completeness
relation}:
\begin{equation} \label{dev2}
 \int \vert \alpha \rangle \langle \alpha \vert \frac{d\bar{\alpha}
  d\alpha}{2\pi \I}=\int e^{-\bar{\alpha} \alpha} e^{\alpha \bar{u}}
  e^{\bar{\alpha } v} \frac{d\bar{\alpha}  d\alpha }{2\pi \I}=e^{\bar{u}v},
\end{equation}
where we used the fact that $e^{\bar{u} v}$ is the kernel
of the identity operator $I$. Next consider a self-adjoint
operator $H_{{\rm para \, }}(g) = \I g(a^{2} - a^{\dagger 2})$, the
generator of parametric amplification\index{parametric
amplification}. In order to have the kernel $U(\bar{u} ,
v)$ of the operator $e^{\I H_{{\rm para \, }}(g)}$, we use
Theorem \ref{thm4.4}, i.e., $B = -\I g$ and $C = 0$, and
$$  {\mathcal A}  = \left( \begin{array}{ccccc} 0 & \I g \\ \I g & 0 \end{array}
\right) , \ \     \left( \begin{array}{ccccc} \Phi &\Psi
\\{} \bar{\Psi }  &\bar{\Phi }  \end{array} \right) = e^{\I t
\mathcal A} = \exp tg \left(
\begin{array}{ccccccc} 0&-1\\{} -1&0\end{array} \right) $$
$$  = \cosh tg  \left( \begin{array}{ccccc} 1&0\\{}
0&1\end{array} \right)  + \sinh tg  \left(
\begin{array}{ccc} 0&-1\\{} -1&0\end{array} \right) =
\left( \begin{array}{ccc} \cosh{}tg&-\sinh{}tg\\{}
-\sinh{}tg&\cosh{}tg\end{array} \right) ,$$
$$ A^{22} = -\Phi ^{-1}\Psi  = \tanh tg, \  A^{21} = \Phi^{-1} = \cosh^{-1} tg , \  A^{11} = \bar{\Psi }  \Phi^{-1} = -\tanh tg.$$
The kernel $U(\bar{u},v)$ of the operator $e^{\I H_{{\rm para \,}}(g)}$ therefore is
$$ U(\bar{u},v) = (\cosh g)^{-1/2} \exp (1/2)\{\tanh g \bar{u}^{2} +
2\cosh^{-1} g \bar{u} v - \tanh g v^{2}\}.$$
In the holomorphic representation, its action on the vacuum state can be
calculated explicitly
$$ e^{\I H_{{\rm para \,}}(g)}\vert 0\rangle = \int U(\bar{u},v) e^{-\bar{v} v}
\frac{d\bar{v}  dv}{2\pi \I}$$
\begin{multline*} = (\cosh g)^{-1/2} \exp (1/2)\{ \tanh g \bar{u}^{2}\}\\ \times \int
 \exp (1/2)\{ 2\cosh ^{-1} g \bar{u}  v - \tanh g v^{2}\} e^{-\bar{v}  v}
 \frac{d\bar{v} dv}{2\pi \I}\end{multline*}
\begin{equation} \label{dev3}
    = (1 - \tanh ^{2} g)^{1/4} \exp (1/2)\{ \tanh g \bar{u}^{2}\}
\end{equation}
where we used the relation (\ref{4.2}).  The state
(\ref{dev3}) is called the squeezed vacuum\index{squeezed
vacuum} with squeezing parameter\index{squeezing parameter}
$g$. \vskip 12pt \noindent Let us consider the
beam splitter\index{beam splitter}.  The generator of the
beam splitter $H_{{\rm bs \, }}(\theta )$ is defined by
$$ H_{{\rm bs \, }}(\theta ) = \I\theta (a_{1}^{\dagger}a_{2} - a_{1}a_{2}^{\dagger}).$$
Then applying the Theorem \ref{thm4.4} again, we get the
kernel $U(\bar{u},v)$ of $e^{\I H_{{\rm bs \,}}(\theta)}$ as follows
$$ B = 0, \  C = \left( \begin{array}{ccccc} 0&\I\theta \\{} -\I\theta &0\end{array} \right), \
 {\mathcal A}= \left( \begin{array}{ccc} -C&0\\{} 0&\bar{C} \end{array} \right) ,$$
$$\left( \begin{array}{ccccc}\Phi &\Psi \\{}\bar{\Psi} &\bar{\Phi}\end{array} \right)= e^{\I
t{\mathcal A} } = \left( \begin{array}{ccc} e^{-\I C}&0\\{} 0&e^{\I t\bar{C} }\end{array} \right) ,$$
$$ e^{-\I tC} = e^{\I t\bar{C}} = \exp t\theta \left( \begin{array}{ccccccc} 0&1\\{} -1&0\end{array} \right)
 = \cos t\theta  \left( \begin{array}{ccccc} 1&0\\{} 0&1\end{array} \right)  + \sin t\theta  \left( \begin{array}{ccc} 0&1\\{} -1&0\end{array} \right) $$
$$ \Phi =\left(\begin{array}{ccc}\cos{}t\theta &\sin{}t\theta \\{} -\sin{}t\theta &\cos{}t\theta \end{array} \right), \ \Psi=0.$$
$$ U(\bar{u}, v) = c \exp \left( \frac{1}{2}(v \  \bar{u}) \left( \begin{array}{ccc} A^{11}&A^{12}\\{}
A^{21}&A^{22}\end{array} \right) \left( \begin{array}{c}v\\{} \bar{u}  \end{array} \right) \right) , \  A^{jk} = {}^{t}A^{kj}$$
$$ A^{22} = A^{11} = 0, \  A^{21} = \Phi ^{-1} = \left( \begin{array}{ccc} \cos{}\theta &-\sin{}\theta
\\{} \sin{}\theta &\cos{}\theta \end{array} \right) , \  c = (\det \Phi e^{\I C})^{-1/2} = 1.$$
The beam splitter for $\theta  = \pi /4$ is called the {\it half-beam splitter\index{half-beam splitter}} whose generator is $H_{{\rm hbs \,} } = \I(\pi /4)(a_{1}^{\dagger}a_{2} -a_{1}a_{2}^{\dagger})$. The half-beam splitter transforms the state $f(\bar{u}_{1}, \bar{u}_{2})$ into
$$ \int U(\bar{u},v) f(\bar{v}_{1}, \bar{v}_{2}) e^{-\bar{v} v} \prod_{j=1}^{2} \frac{d\bar{v} _{j} dv_{j}}{2\pi \I}$$
$$  \int \exp \frac{1}{\sqrt{2}}\{(\bar{u}_{1} + \bar{u}_{2})v_{1}+(-\bar{u}_{1} + \bar{u}_{2})v_{2}\} f(\bar{v}_{1}, \bar{v}_{2}) e^{-\bar{v} v} \prod _{j=1}^{2} \frac{d\bar{v} _{j} dv_{j}}{2\pi \I}$$
\begin{equation} \label{dev4}
    = f((\bar{u} _{1} + \bar{u} _{2})/\sqrt{2}, (-\bar{u} _{1} +\bar{u} _{2})/\sqrt{2}).
\end{equation}

\nopagebreak
\subsection{Balanced homodyne detection} We
calculate the mean value of the difference of photon beams
$N_{1} - N_{2} = a_{1}^{\dagger}a_{1} - a_{2}^{\dagger}a_{2}$ for the
state $e^{-\I H_{hbs}}\vert \psi_{1}\rangle \otimes \vert
\psi_{2}\rangle $, the state after passing through the
half-beam splitter.  Since
$$ e^{\I H_{hbs}}a_{1}e^{-\I H_{hbs}}=(a_{1} + a_{2})/\sqrt{2}, \  e^{\I H_{hbs}}a_{2}e^{-\I H_{hbs}} = (-a_{1} + a_{2})/\sqrt{2},$$
$$ \langle \psi_{1}\vert \otimes \langle \psi_{2}\vert e^{\I H_{hbs}}(N_{1} - N_{2})e^{-\I H_{hbs}}\vert \psi _{1}\rangle  \otimes \vert \psi _{2}\rangle $$
$$= \langle \psi _{1}\vert \otimes \langle \psi _{2}\vert (a_{1}^{\dagger}a_{2} + a_{1}a_{2}^{\dagger})\vert \psi _{1}\rangle  \otimes \vert \psi _{2}\rangle .$$
If $\vert \psi _{2}\rangle $ is a coherent state $\vert
\alpha_{2}\rangle $, then this identity is continued by
$$= \langle \psi_{1}\vert \otimes \langle \alpha_{2}\vert (a_{1}^{\dagger}a_{2} + a_{1}a_{2}^{\dagger})\vert \psi_{1}\rangle \otimes \vert \alpha_{2}\rangle
 =  \langle \psi_{1} \vert a_{1}^{\dagger} \vert \psi_{1}\rangle \alpha_{2} + \langle \psi_{1} \vert a_{1} \vert \psi _{1}\rangle \bar{\alpha}_{2}$$
$$ = \langle \psi _{1}\vert (q_{1}-\I p_{1})/\sqrt{2}\vert \psi _{1}\rangle \vert \alpha _{2}\vert e^{\I\theta_{2}} + \langle \psi_{1}\vert(q_{1}+\I p_{1})/\sqrt{2}\vert \psi_{1}\rangle \vert \alpha _{2}\vert e^{-\I\theta _{2}}$$
$$ = \sqrt{2}\langle \psi _{1}\vert q_{1}\vert \psi _{1}\rangle \vert \alpha _{2}\vert  \cos \theta _{2} + \sqrt{2}\langle \psi _{1}\vert p_{1}\vert \psi _{1}\rangle \vert \alpha _{2}\vert \sin \theta _{2}$$
$$= \sqrt{2}\vert \alpha _{2}\vert \langle \psi _{1}\vert (q_{1} \cos \theta _{2} + p_{1} \sin \theta _{2})\vert \psi _{1}\rangle .$$
Thus the balanced homodyne detection\index{balanced
homodyne detection} measures $\sqrt{2} \vert \alpha _{2}
\vert $ times of $q_{1}$ and $p_{2}$ according to the phase
of the coherent beam $\alpha _{2}$.\newline The
displacement operator $D(-\alpha  )$ is realized by
modulators and the beam splitter.  For $\vert  \phi \rangle
\in  {\mathcal H} _{2}$ and $\vert \beta \rangle \in
{\mathcal H} _{3}$, we send $\vert \phi \rangle \otimes
\vert \beta \rangle  $ through a beam splitter $B(\theta ) =
e^{\I H_{bs}(\theta )}$.  Let
$$ \vert \phi \rangle  \otimes \vert \beta \rangle = \phi (\bar{u}_{2}) e^{-\vert \beta \vert ^{2}/2} e^{\beta \bar{u} _{3}}.$$
Then
$$ B(\theta)(\vert \phi \rangle  \otimes \vert \beta \rangle)=B(\theta) \phi (\bar{u}_{2}) e^{-\vert \beta \vert ^{2}/2} e^{\beta \bar{u}_{3}}$$
$$= \phi (\bar{u}_{2} \cos \theta  + \bar{u}_{3} \sin \theta ) e^{-\vert \beta \vert ^{2}/2} e^{\beta (-\bar{u}  _{2} \sin \theta  + \bar{u}  _{3} \cos \theta )}$$
$$= \phi (\bar{u}  _{2} \cos \theta  + \bar{u}  _{3} \sin \theta ) e^{-\vert \beta \vert ^{2}/2} e^{-\beta \bar{u}_{2} \sin \theta }e^{\beta \bar{u}  _{3} \cos \theta }.$$
$$ (I\otimes \langle \beta \vert )B(\theta )\vert \phi \rangle  \otimes \vert \beta \rangle  $$
$$ = \int e^{-\vert \beta \vert ^{2}} e^{\beta u_{3}} \phi (\bar{u}  _{2} \cos \theta  + \bar{u}  _{3} \sin \theta )e^{-\beta \bar{u}_{2} \sin \theta }e^{\beta \bar{u}_{3} \cos \theta} e^{-\bar{u}_{3}u_{3}} \frac{d\bar{u}_{3}du_{3}}{2\pi \I}$$
$$= e^{-\vert \beta \vert^{2}} e^{-\beta \bar{u}_{2} \sin \theta } \int e^{\bar{\beta} u_{3}}e^{\beta \bar{u}  _{3} \cos \theta} \phi (\bar{u}_{2} \cos \theta + \bar{u}_{3} \sin \theta) \frac{d\bar{u}_{3}du_{3}}{2\pi \I}$$
$$ = e^{-\vert \beta \vert^{2}} e^{-\beta \bar{u}_{2} \sin \theta} e^{\beta \bar{\beta}\cos \theta}\phi (\bar{u} _{2} \cos \theta  + \bar{\beta}\sin \theta ).$$
Let $\theta  \rightarrow  0$ and $\beta  \rightarrow \infty $ such that $\beta  \sin \theta  \rightarrow  \alpha $.  Then
$$ \vert \beta \vert^{2}(1 - \cos \theta) \rightarrow  \vert \alpha \vert^{2}/2$$
and
$$ e^{-\vert \beta \vert^{2}} e^{-\beta \bar{u}_{2} \sin \theta} e^{\beta \bar{\beta} \cos \theta } \phi (\bar{u}_{2} \cos \theta + \bar{\beta} \sin \theta )$$
$$ \rightarrow  e^{-\vert \alpha \vert /2} e^{-\alpha \bar{u}_{2}} \phi (\bar{u}_{2} + \bar{\alpha}) = D(\alpha )\vert \phi \rangle .$$

\section{Teleportation of continuous quantum variables}
The proposal for continuous variable quantum teleportation was
first made by Vaidman [\cite{Va94}] in 1994, and then by
Braunstein and Kimble [\cite{BK98}] in 1998, and experimentally
demonstrated by the Caltech group, Furusawa et al. [\cite{Fu98}]
in 1998. Now we present the teleportation of continuous quantum
variables in a parallel way as the teleportation of qubits in
Section 2. The numbering $(1), \ldots , (4)$ corresponds to that
in Section 2. Figure 2  is similar to the Figure 1 and these
figures illustrate the correspondence of the two cases of
teleportation.
\begin{figure}[!h]
  \centering
 \includegraphics[width=12cm]{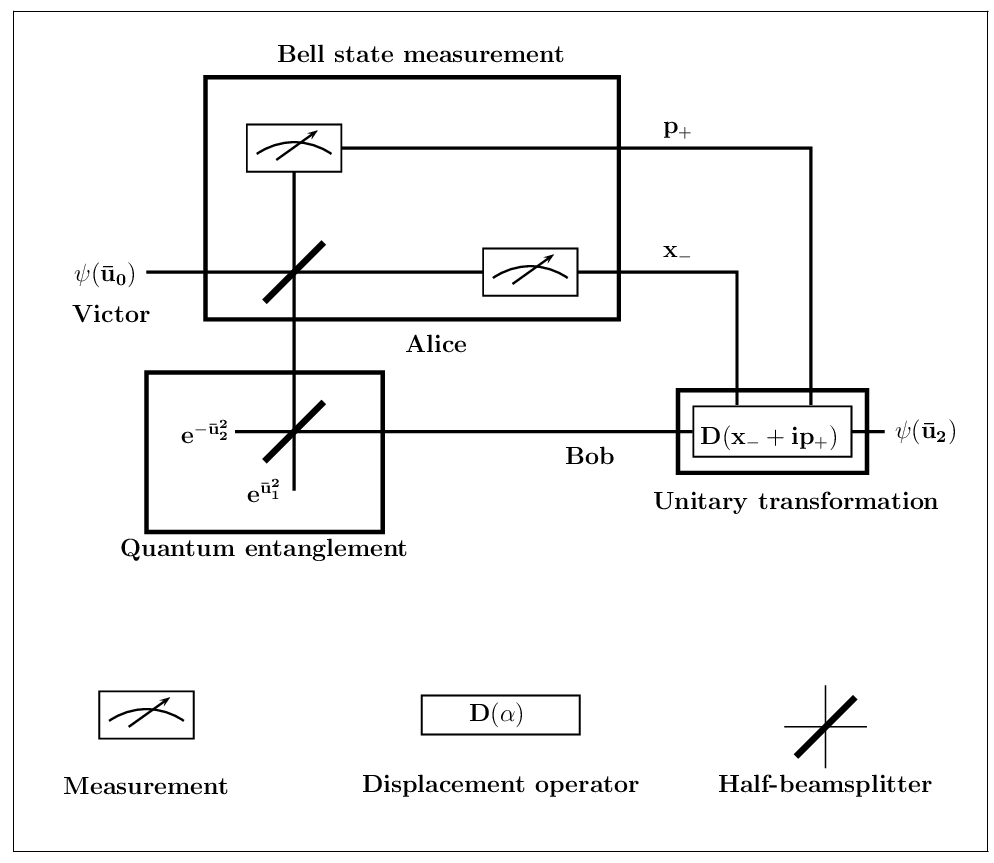}
    \caption{Quantum circuit of continuous variable
 teleportation}
    \label{fig:teleportcont}
\end{figure}
\begin{enumerate}
\item[(1)] First,  an entangled state as a counterpart to an
EPR pair\index{EPR pair} is produced. We prepare such a state using
parametric amplification\index{parametric amplification}
and a beam splitter. By parametric amplification we create a
pair of squeezed vacuum\index{squeezed vacuum} states
$U_{{\rm para \, }}(g)\vert 0\rangle  = e^{\I H_{{\rm para \,
}}(g)}\vert 0\rangle $ and $U_{{\rm para \, }}(-g)\vert
0\rangle $.  In the holomorphic representation one has (see
(\ref{dev2})),
$$ (1 - q^{2})^{-1/4} U_{{\rm para \, }}(g)\vert 0\rangle  = e^{q\bar{u}  _{1}^{2}/2},$$
$$(1 - q^{2})^{-1/4}U_{{\rm para \, }}(-g)\vert 0\rangle  = e^{-q\bar{u}  _{2}^{2}/2}, \  q = \tanh g.$$
Then we send these states through the half-beam splitter
(see (\ref{dev4})). Now Alice and Bob share the state
($\bar{u_{1}}$ is Alice's variable and $\bar{u}_{2}$ is Bob's variable)
$$ \vert \Psi_{0}\rangle= e^{q(\bar{u}_{1}+\bar{u}_{2})^{2}/4}e^{-q(-\bar{u}_{1}+\bar{u}_{2})^{2}/4} = e^{q\bar{u}_{1}\bar{u}_{2}}, \  q = \tanh g.$$
\indent Victor gives Alice a state $\psi (\bar{u} _{0})$ to
send to Bob.  The state of the total system is
$$ \vert \Psi_{1}\rangle=\psi (\bar{u}_{0})e^{q\bar{u}_{1}\bar{u}_{2}} = \vert \psi \rangle_{0} \otimes  \sum_{n=0}^{\infty } q^{n}\vert n\rangle _{1} \otimes \vert n\rangle _{2}, \  q = \tanh g.$$
This state corresponds to the state
$$ \vert \psi _{0}\rangle  = \vert \psi \rangle _{V} \otimes \frac{1}{\sqrt{2}}(\vert 0\rangle _{A}\vert 0\rangle _{B} + \vert 1\rangle_{A}\vert 1\rangle _{B})$$
of the qubit case.
The generalized state (for some background information on generalized states see the first part of the Appendix)
$$ \pi^{-1/2} e^{\bar{v}_{0}\bar{v_{1}}} = \pi ^{-1/2} \sum _{n=0}^{\infty} \frac{(\bar{v}_{0}\bar{v}  _{1})^{n}}{n!}= \pi^{-1/2} \sum_{n=0}^{\infty} \vert n\rangle_{0} \otimes \vert n\rangle_{1}
$$
corresponds to the Bell state $\vert \beta_{00}\rangle= 2^{-1/2} \sum_{n=0}^{1} \vert n\rangle_{0} \otimes \vert n\rangle_{1}$ of Section 2.  The other Bell states $\vert \beta _{ij}\rangle $ correspond to the states
$$ \sum_{n=0}^{\infty}(D(\alpha )\vert n\rangle_{0})\otimes \vert n\rangle_{1} = e^{-\vert \alpha \vert ^{2}/2} e^{(\bar{v}_{0} - \bar{\alpha }) \bar{v}_{1}} e^{\alpha \bar{v}_{0}} = e^{-\vert \alpha \vert ^{2}/2} e^{\bar{v}  _{0}\bar{v} _{1}} e^{-\bar{\alpha }  \bar{v} _{1}} e^{\alpha \bar{v}_{0}} $$ (see (\ref{dev1})).

In the Appendix we show that
\begin{equation} \label{5.1}
      \{ \pi ^{-1/2} \sum _{n=0}^{\infty } (D(\alpha )\vert
n\rangle _{0})\otimes \vert n\rangle _{1}; \alpha  \in
\mathbb C\}, \  \alpha = x_{-} + i p_{+}
\end{equation}
is the generalized Bell basis\index{Bell basis}, i.e., a complete orthonormal system in our Hilbert space.

Then we send this state through the half-beam splitter realized as the unitary operator $e^{iH_{{\rm hbs \, }}}$, and  obtain (see (\ref{dev4}))
$$  e^{-(x_{-}^{2}+p_{+}^{2})/2} e^{-\bar{u}_{0}^{2}/2} e^{\bar{u}_{1}^{2}/2} e^{\sqrt{2}x_{-}\bar{u}  _{0}}e^{\I\sqrt{2}p_{+}\bar{u}_{1}}=\pi^{1/2} \vert x_{-}\rangle  \otimes \vert p_{+}\rangle ,$$ where
$$ \vert x_{-}\rangle=\pi^{-1/4} e^{-x_{-}^{2}/2} e^{-\bar{u}_{0}^{2}/2}e^{\sqrt{2}x_{-}\bar{u}_{0}}$$
is the generalized eigen-state of the operator
$$ x_{0} = \frac{1}{\sqrt{2}} \left(\frac{\partial}{\partial \bar{u}_{0}} + \bar{u}_{0}\right) $$
with eigen-value $x_{-}$, and
$$\vert p_{+}\rangle=\pi^{-1/4} e^{-p_{+}^{2}/2} e^{\bar{u}_{1}^{2}/2}e^{i\sqrt{2}p_{+}\bar{u}_{1}}$$
is the generalized eigen-state of the operator
$$ p_{1}=\frac{1}{\sqrt{2}i} \left(\frac{\partial}{\partial \bar{u}_{1}} - \bar{u}_{1}\right) $$
with eigen-value $p_{+}$. In the Appendix  the orthogonality relations
\begin{equation}\label{orth-rel}
\begin{aligned}
& \langle x_{-}^{\prime}\vert x_{-}\rangle = \delta(x_{-}^{\prime}-x_{-}),
&\langle p_{+}^{\prime }\vert p_{+}\rangle=\delta (p_{+}^{\prime}-p_{+}).
\end{aligned}
\end{equation}
are shown. Thus we have
\begin{equation} \label{5.2}
   e^{\I H_{{\rm hbs \, }}} \sum _{n=0}^{\infty } (D(\alpha )\vert
n\rangle_{0}) \otimes \vert n\rangle_{1}=\pi ^{1/2}\vert x_{-}\rangle\otimes \vert p_{+}\rangle
\end{equation}
and
$$\pi^{-1}(\sum_{m=0}^{\infty}{}_{0}\langle m \vert(D(\alpha^{\prime})^{\dagger}\otimes{}_{1}\langle m \vert) ( \sum _{n=0}^{\infty } (D(\alpha )\vert n\rangle_{0})\otimes \vert n\rangle_{1} )$$
$$= (\langle x_{-}^{\prime }\vert \otimes \langle p_{+}^{\prime }\vert),(\vert x_{-}\rangle \otimes \vert p_{+}\rangle)=\delta (x_{-}^{\prime} - x_{-})\delta (p_{+}^{\prime}-p_{+}).$$
\vskip 12pt \noindent The state $\vert \psi \rangle_{V}\otimes \vert \beta_{00}\rangle_{AB}$ corresponds to
$$ \vert \Phi_{1}\rangle=\pi^{-1/2}\vert \psi \rangle_{0} \otimes \sum_{n=0}^{\infty} q^{n} \vert n\rangle_{1} \otimes \vert n\rangle_{2}.$$
The relation
$$I=\pi^{-1} \int dx_{-}dp_{+} \sum_{m=0}^{\infty}(D(\alpha )\vert m\rangle_{0})\otimes \vert m\rangle _{1} \sum _{k=0}^{\infty}{}_{0}\langle k\vert D(\alpha)^{\dagger} \otimes{}_{1}\langle k\vert$$
implies for $\alpha=x_{-} + \I p_{+}$ that
$$\vert \Phi_{1}\rangle=\pi^{-3/2} \int dx_{-}dp_{+} \sum _{m=0}^{\infty}(D(\alpha )\vert m\rangle _{0})\otimes \vert m\rangle_{1}$$
$$\otimes \sum_{k=0}^{\infty} \sum_{n=0}^{\infty} {}_{0}\langle k\vert D(\alpha )^{\dagger}\vert \psi \rangle_{0} q^{n} {}_{1}\langle k\vert n\rangle_{1}\vert n\rangle_{2}$$
$$=\pi^{-3/2}\int dx_{-}dp_{+}\sum_{m=0}^{\infty}(D(\alpha )\vert m\rangle_{0})\otimes \vert m\rangle_{1}
\otimes \sum_{n=0}^{\infty} q^{n}{}_{0}\langle n\vert D(\alpha)^{\dagger}\vert \psi \rangle_{0}\vert n\rangle_{2}.$$
This is the expansion of $\vert \Phi _{1}\rangle $ with respect to the generalized Bell basis (\ref{5.1}).
\item[(2)] Then Alice performs the (generalized) Bell-state
measurement\index{Bell-state measurement}, and when some Bell state $\pi^{-1/2} \sum_{n=0}^{\infty} (D(\alpha)\vert n\rangle_{0})\otimes \vert n\rangle_{1}$ is
chosen, then the total system is reduced to
$$=\pi^{-1/2} \sum_{m=0}^{\infty}(D(\alpha)\vert m\rangle_{0})\otimes \vert m\rangle_{1} \otimes \sum _{n=0}^{\infty} q^{n}{}_{0}\langle n\vert D(\alpha)^{\dagger}\vert \psi \rangle_{0}\vert n\rangle_{2}.$$ If $q = 1$, the above state is
$$\sum_{m=0}^{\infty}(D(\alpha)\vert m\rangle_{0})\otimes \vert m\rangle_{1}\otimes D(\alpha)^{\dagger} \vert \psi \rangle_{2}.$$
Thus the Bob's state is reduced to $D(\alpha)^{\dagger} \vert \psi \rangle_{2}$ though he can not know this.
\item[(3)] Alice sends the classical information $\alpha = x_{-} +\I p_{+}$ to Bob. Then Bob sends Bob's state $D(\alpha)^{\dagger} \psi $ through $D(\alpha)$ according to the result $\alpha
= x_{-} + \I p_{+}$, obtaining $\psi_{{\rm out}}= D(\alpha)D(\alpha)^{\dagger} \psi =\psi $. Alice sends two real numbers $(x_{-}, p_{+})$ to Bob, and Bob gets a state $\psi= \sum_{n=0}^{\infty} a_{n} \vert n\rangle $ in an infinite dimensional Hilbert space.
\item[(4)] The Bell-state measurement (2) is done by sending the Bell basis through the half-beam splitter obtaining the canonical basis $\{ \vert x_{-}\rangle \otimes \vert p_{+}\rangle ;
\alpha  = x_{-} + \I p_{+} \in  \mathbb C\} $ of (\ref{5.2})
and by the measurement of $x_{0}$ and $p_{1}$ using balanced homodyne detection.

Since we cannot generate an EPR pair $e^{\bar{u}_{1}\bar{u}_{2}}$ with an
infinite squeezing parameter $g = \infty $, the ideal $q = 1$ case of the squeezed state $e^{q\bar{u}  _{1}\bar{u}_{2}}$, $q = \tanh g < 1$,  we cannot have  complete teleportation $\psi_{\mathrm{out}}=\psi(= \psi _{\mathrm{in}})$.  We have to measure the distance between $\psi $ and $\psi_{\mathrm{out}}$.
For density operators $\rho $ (positive operator with ${\rm tr \,} \rho = 1$) and $\sigma $, the
fidelity\index{fidelity} $ F(\rho , \sigma )$, given by
$$ F(\rho,\sigma)={\rm tr \,}\sqrt{\rho^{1/2}\sigma \rho^{1/2}}$$
is an accepted measure for such a distance. Let $\vert \psi \rangle = \vert \gamma \rangle $ be a coherent state, and
$$ \vert \phi \rangle=\sum _{n=0}^{\infty} q^{n} D(\alpha)\vert n\rangle \langle n\vert D(\alpha )^{\dagger}\vert \gamma \rangle .$$
For $\rho = \vert \gamma \rangle \langle \gamma \vert $
and $\sigma  = \vert \phi \rangle \langle \phi \vert $,
$$  \sqrt{\rho^{1/2} \sigma  \rho ^{1/2}} = \sqrt{\rho  \sigma \rho} = \sqrt{\vert \gamma \rangle \langle \gamma \vert \phi \rangle \langle \phi \vert \gamma \rangle \langle \gamma \vert}=\vert \langle \gamma \vert \phi \rangle \vert \vert \gamma \rangle \langle \gamma \vert $$
and the fidelity $F(\rho,\sigma)$ equals $\vert \langle \gamma \vert \phi \rangle \vert $.  The inner product $\langle \gamma \vert \phi \rangle $ is easily calculated, using  relation (\ref{4.1}):
$$\langle \gamma \vert \phi \rangle = \sum_{n=0}^{\infty} q^{n}\langle \gamma \vert D(\alpha)\vert n\rangle \langle n\vert D(\alpha)^{\dagger}\vert \gamma \rangle
 = \int \int e^{-\vert \alpha \vert^{2}/2} e^{-\bar{\alpha} u}
 e^{-\vert \gamma \vert^{2}/2}e^{\bar{\gamma}(\bar{u} + \bar{\alpha})} e^{q\bar{u}v}$$
$$\times e^{-\vert \alpha \vert^{2}/2} e^{-\alpha \bar{v}}e^{-\vert
\gamma\vert^{2}/2}e^{\gamma(\bar{v}+\bar{\alpha})}e^{-\bar{u} u}e^{-\bar{v}v} \frac{d\bar{u} du}{2\pi \I}\frac{d\bar{v} dv}{2\pi \I}$$
$$ = \exp(1-q)\{\alpha \bar{\gamma}+\bar{\alpha}\gamma-\vert \alpha
  \vert^{2} - \vert \gamma \vert^{2}\} .$$
  \end{enumerate}

\section{Experiment, Controversy and Locality}
\subsection{The Experiment} Here, we briefly explain the experiment
by the Caltech group, Furusawa et al. [\cite{Fu98}].  The
teleported state is not an optical beam itself but a modulation
sideband of a bright optical beam generated by an electro-optical
modulator, because the frequency of the optical beam is too high
($\omega /\pi $ = 300THz, wavelength 1000 nm) to handle directly.
So, the optical beam is treated only as a carrier and the quantum
states are discussed using sideband frequency. The light from a
single-frequency titanium sapphire (TiAl$_{2}$O$_{3}$) laser at
860 nm (frequency $\omega _{L}$) serves as the primary source for
all fields in the experiment.  90 percent of the laser output with
frequency $\omega _{L}$ is directed to a frequency-doubling cavity
to generate blue light at $2\omega _{L}$.  This output then splits
into two beams that serve as harmonic pumps for parametric
down-conversion, $2\omega _{L} \rightarrow  \omega _{L} \pm \Omega
$, within the optical parametric oscillator (OPO).
\begin{figure}[!h]\begin{center}
\includegraphics[width=12cm]{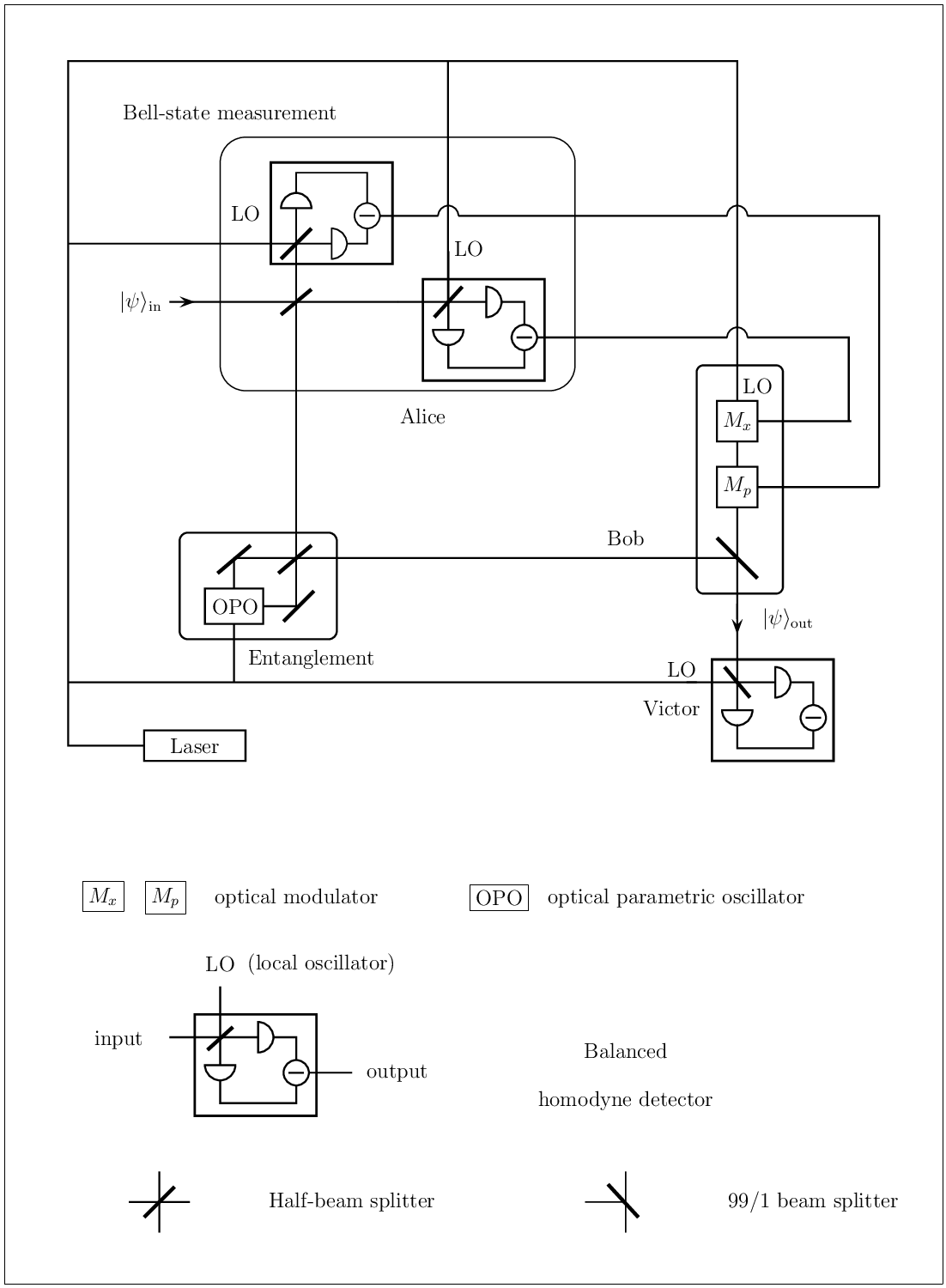}
\caption{Experimental scheme for quantum teleportation performed
by Caltech group. The laser field is shared by all parties.}
\end{center}
\end{figure}
\begin{enumerate}
\item[(1)]   Thus the two independent squeezed beams which are supposed to be represented by
$e^{q\bar{u}_{1}^{2}}$ and $e^{-q\bar{u}_{2}^{2}}$ at 860 nm are
generated by optical parametric oscillators\index{parametric
oscillator} (amplifiers), where $q = \tanh g$ and $g$ is called
the squeezing parameter. These beams are sent through a
beamsplitter obtaining the EPR beam $e^{q\bar{u} _{1}\bar{u}
_{2}}$ and the outcomes are sent to Alice (variable $\bar{u}
_{1}$) and Bob (variable $\bar{u} _{2}$).\newline \indent Victor
generates a coherent sideband $\vert \psi \rangle = e^{-\vert
\alpha \vert /2}e^{\alpha \bar{u} _{0}}$ at frequency $(\omega \pm
\Omega )/2\pi $ with $\Omega /2\pi $ = 2.9 MHz by an
electro-optical modulator, and sends it to Alice.

 If $e^{\I t\omega }$ represents the carrier and $\alpha e^{\pm \I t
\Omega }$ the modulation, then the total system is represented by
$$ \alpha e^{\pm \I t \Omega} e^{\I t\omega}= \alpha e^{\I(\omega \pm \Omega)} .$$
\item[(2)]
   \begin{enumerate}
\item
 Alice sends two beams, $e^{-\vert \alpha \vert
/2}e^{\alpha \bar{u}_{0}}$ and her EPR beam, through a
half-beamsplitter $e^{\I H_{\mathrm{hbs}}}$ discussed in Section 5, obtaining
$$ e^{-\vert \alpha \vert /2}e^{\alpha (\bar{u}_{0}+\bar{u}_{1})\sqrt{2}}e^{q(-\bar{u}_{0} + \bar{u}  _{1})\bar{u} _{2}/\sqrt{2}}.$$
\item   Alice uses two sets of balanced homodyne
detectors\index{balanced homodyne detector} $(D_{x_{0}},
D_{p_{1}})$ discussed in Section 5 to make a joint measurement of
the amplitude $x_{0}$ and $p_{1}$ discussed in Section 6.
\end{enumerate}
\item[(3)]   Alice sends the outcomes $(x_{-}, p_{+})$ of the detectors
to Bob.
\item[(4)]After receiving this classical information from
Alice, Bob is able to construct the teleported state $\rho
_{\mathrm{out}}$.  Bob generates a sideband beam at frequency $(\omega \pm
\Omega)/2\pi $ by two electro-optical modulators $M_{x}$
(amplitude modulator) and $M_{p}$ (phase modulator) with suitable
complex amplitude $\vert \beta \rangle= e^{-\vert \beta \vert /2}
e^{\beta \bar{u}_{3}}$. Then he sends $\vert \beta \rangle $ and
his EPR beam through a beamsplitter of refractivity 0.99 obtaining
the state $\rho_{\mathrm{out}}$.
\item[(5)] Victor detects the state $\rho_{out}$ by his own balanced homodyne detector $D_{V}$ and
compares it with his original state $\vert \psi \rangle $.
\end{enumerate}
\noindent
 The result obtained by Furusawa, A., et al. is that the fidelity
$F(\vert \psi \rangle, \rho_{\mathrm{out}})$ of the states $\vert \psi
\rangle $ and $\rho _{\mathrm{out}}$ is
$$ F(\vert \psi \rangle, \rho_{\mathrm{out}}) = 0.58 \pm  0.02.$$
Bowen, W.P., et al. [\cite{Bo03}] had the fidelity of $0.64 \pm
0.02$ and Takei, N., et al. [\cite{Ta05}] achieved the fidelity of
$0.70 \pm 0.02$.

\nopagebreak
\subsection{Controversy}  After the
experimental demonstration of continuous variable quantum
teleportation (CVQT) was reported [\cite{Fu98}], there was a
controversy over its validity on the ground of intrinsic phase
indeterminacy of the laser field [\cite{RS01}].  The laser field
is often assumed to be a coherent state having a fixed phase, but
[\cite{RS01}] shows that the steady-state solution of the master
equation in the quantum theory of the laser shows that the phase
of the laser field inside the cavity is completely unknown and
genuine CVQT cannot be achieved using conventional lase sources,
due to an absence of optical coherence.  Furthermore, the same
laser source is used for (i) producing Victor's state for
teleportation, (ii) pumping the nonlinear crystal that produces a
two-mode squeezed light field serving as the shared EPR state,
(iii) supplying local oscillator (LO) fields for both of Alice's
homodyne measurements, and (iv) providing Bob with a coherent
field to mix with his portion of the EPR beam to reconstruct
Victor's state.\newline The ideal scheme for CVQT [\cite{BK98}] is
explained in Fig. 2 and an explanation of the experimental
simulation of CVQT [\cite{Fu98}] is provided in Fig. 3. \newline
For the claim [\cite{RS01}], there appeared a counterargument
[\cite{EF02}] which says that the standard description of the
laser field used in [\cite{RS01}] is insufficient to understand
CVQT with a laser.  They found that the laser light has the random
phase only in the cavity, and outside the cavity, it has similar
phase character as the coherent state has, and therefore a
conventional laser can be used for CVQT.\newline Furthermore,
[\cite{Fu03}] says that the laser field outside the cavity is a
mixed state whose phase is completely unknown, but CVQT with a
laser is valid only if  the unknown phase of the laser field is
shared among sender's LOs, the EPR state, and receiver's LO.
\subsection{Locality} Let $f_{i}(x)$ be two functions square integrable functions
on Euclidean space, i.e., elements in $L^{2}(\mathbb{R}^{3})$, and suppose that the support of the function
$f_{i}$ is contained in a bounded set $O_{i}$ of $\mathbb{R}^{3}$.  Then for a
system in the state represented by the function $f_{i}(x)$, the
observable $x$ (the position of a particle) is always observed in
the set $O_{i}$.  In such situation, we often say that the state
$f_{i}$ is localized in $O_{i}$.  The state of two particles at $x_{1}$
and $x_{2}$ which are separated by a long distance is expressed by
the tensor product $f(x_{1}, x_{2}) = f_{1}(x_{1})f_{2}(x_{2})$ of
two functions $f_{i}$ ($i = 1, 2$) whose supports $O_{i}$ ($i = 1,
2$) are separated by a long distance.

Ideal quantum teleportation can be considered as the following process: The information contained
in a quantum state $\vert \psi \rangle_{{\rm in \,}}$ (which is localized
in a region $O_{1}$) is sent to another region $O_{2}$ separated
from $O_{1}$ by a long distance and produces a state $\vert \psi
\rangle _{{\rm out \,}}$ (which is localized in $O_{2}$) and
contains all the information of $\vert \psi \rangle _{{\rm in \,
}}$.

In the framework of teleportation of qubits in
Section 2 where a finite dimensional Hilbert space is used, it
seems difficult to define a state localized in some bounded
region.  In the framework of CVQT in Section 6, though an infinite
dimensional Hilbert space is used, we manipulate only
monochromatic laser beams.  When the frequency of the laser beam is
determined, the location of the laser beam is completely unknown.
Up to now, one has considered  quantum teleportation without the notion
of locality. In a precise formulation of quantum teleportation which takes locality into account,
a quantum theory of infinitely many degrees of freedom might be needed, e.g.,  a quantum field
theory might be necessary.

 It seems that the theory of lasers is not satisfactory in its application to quantum teleportation,
and therefore, many controversies appear around CVQT which uses laser beams.
In this context to, a theory of quantum teleportation based on quantum field theory might be preferable.

\begin{appendix}
\section{}
\subsection{Generalized vectors and generalized eigen-states}
The theory of the generalized vectors is developed in
[\cite{GV64}]. Let ${\mathcal S}(\mathbb{R})$ the Schwartz space
of fast decaying infinitely often differentiable functions on the
real line $\mathbb{R}$ and $L^{2}(\mathbb{R})$ the Hilbert space
of square integrable functions on $\mathbb{R}$. Then
$${\mathcal S}(\mathbb{R}) \subset  L^{2}(\mathbb{R}) \subset
{\mathcal S}^{\prime }(\mathbb{R})$$ is a rigged Hilbert space
(see [\cite{GV64}]). Here ${\mathcal S}^{\prime} (\mathbb{R})$
denotes the topological dual of ${\mathcal S}(\mathbb{R})$.

 The elements
$F$ of ${\mathcal S}^{\prime }$ are called  generalized vectors.
Suppose that an operator $A$ on $L^{2}(\mathbb{R})$ maps ${\mathcal
S}(\mathbb{R})$ into ${\mathcal S} (\mathbb{R})$. Then a generalized
vector $F_{\lambda } \in {\mathcal S} ^{\prime }(\mathbb{R})$ is
called a generalized eigen-vector of $A$ corresponding to an
eigen-value $\lambda $, if
$$  F_{\lambda}(A\phi) = \lambda F_{\lambda}(\phi)$$
holds for every $\phi \in {\mathcal S}(\mathbb{R})$.\newline Let
$$ E_{\lambda} = \{ F \in {\mathcal S}^{\prime}(\mathbb{R}); F(A\phi)=
\lambda F(\phi)\} $$ be the space of generalized eigen-vectors of
$A$ corresponding to the eigen-value $\lambda $.  We associate with
each element $\phi \in  {\mathcal S}(\mathbb{R})$ and each number
$\lambda $ a linear functional $\tilde{\phi}_{\lambda}$ on
$E_{\lambda}$, taking the value $F_{\lambda}(\phi)$ on the element
$F_{\lambda }$ of $E_{\lambda}$.  We call the correspondence $\phi
\rightarrow \tilde{\phi } _{\lambda }$ the spectral decomposition of
the element $\phi$ corresponding to the operator $A$.\newline If
$\tilde{\phi}_{\lambda} \equiv 0$ implies $\phi = 0$, then we say
that the set of generalized eigen-vectors of the operator $A$ is
complete.\newline Let $p = -id/dx$ be the self-adjoint generator of
translations in $L^{2}(\mathbb{R})$ (more accurately $p$ is the
self-adjoint realization of the differential operator $-id/dx$ on
${\mathcal S}(\mathbb{R})$).  Then the identity
$$ (-id/dx) e^{i\lambda x} = \lambda e^{i\lambda x}, \  \lambda  \in
\mathbb{R}$$ shows that $e^{i\lambda x}$ is an eigen-function of the
operator $p$ corresponding to the eigen-value $\lambda $.  The
function $e^{i\lambda x}$ does not belong to $L^{2}(\mathbb{R})$,
but belongs to ${\mathcal S}^{\prime}(\mathbb{R}) \ni  F_{\lambda }
= e^{i\lambda x}$.  The relation
$$ F_{\lambda}(p\phi) = \int \overline{e^{i\lambda x}} (-id/dx)
\phi(x)dx = \int \overline{(-id/dx) e^{i\lambda x}} \phi(x) dx$$
$$ = \lambda \int \overline{e^{i\lambda x}} \phi(x)dx = \lambda
  F_{\lambda}(\phi )$$
shows that $F_{\lambda} = e^{i\lambda x}$ is a generalized eigen-
vector of $p$ corresponding to the eigen-value $\lambda $.  The
spectral decomposition $\tilde{\phi}_{\lambda}$ of $\phi$ is the
Fourier transformation of $\phi $.
$$ \tilde{\phi}_{\lambda} = F_{\lambda}(\phi ) = \int \overline{e^{i\lambda x}}
 \phi(x) dx = \int e^{-i\lambda x} \phi(x)dx = \tilde{\phi}(\lambda).$$
The Fourier inversion formula
$$\phi(x) = \frac{1}{2\pi} \int_{\mathbb{R}}e^{i\lambda x} \tilde{\phi}
 (\lambda) d\lambda $$
shows that $\tilde{\phi}_{\lambda} \equiv 0$ implies $\phi =0$, that
is, $\{ e^{i\lambda x}; \lambda  \in  \mathbb{R}\} $ is a complete
set of generalized eigen-vectors of $p$.
\newline Though it is not possible to find the value of the
momentum $p$ (continuous variable) to be precisely $\lambda $, it
is convenient to say that if the value $\lambda $ of the momentum
$p$ is found for the state $\vert \phi  \rangle  = \phi (x)$, the
state after the measurement is $\vert \lambda  \rangle \langle
\lambda \vert \phi  \rangle  = e^{i\lambda x} \tilde{\phi }
(\lambda )/2\pi $ in the same way as the discrete variables, where
$\vert \lambda \rangle  = e^{i\lambda x}/\sqrt{2\pi }$ is a
generalized vector.\newline Actually, we can only say that
$\lambda $ is contained in the interval $[a, b]$.  In that case,
using the above generalized eigen-states, the state after the
measurement is $$\int _{a}^{b} d\lambda  \vert \lambda \rangle
\langle \lambda \vert \phi \rangle /\sqrt{\int _{a}^{b} d\lambda
\vert \langle \lambda \vert \phi  \rangle \vert ^{2} }$$. Thus we
have the so-called projective measurement, whose name comes from
the fact that $P = \int _{a}^{b} d\lambda  \vert \lambda  \rangle
\langle \lambda \vert $ is a projection.  If we think that the
edge of $[a, b]$ is not sharp, then we take $M = \int _{a}^{b}
d\lambda  \chi (\lambda ) \vert \lambda  \rangle \langle \lambda
\vert $ as a measurement operator, where $\chi (\lambda )$ is a
$C^{\infty }$-function with the support contained in $[a-\epsilon
, b+\epsilon ]$ and $\chi (\lambda ) = 1$ for $\lambda  \in
[a+\epsilon , b-\epsilon ]$ and $\epsilon  > 0$. Then we have the
following generalized measurement (see [\cite{NC00}]).  \newline
The state after the measurement is $$M\vert \phi \rangle
/\sqrt{\langle \phi \vert M^{*}M\vert \phi  \rangle }.$$
\newline

If we introduce the operators
$$a = \frac{1}{\sqrt{2}} \left( x + \frac{d}{dx}\right), \  a^{*}
= \frac{1}{\sqrt{2}} \left( x - \frac{d}{dx}\right),$$  the topology
of ${\mathcal S}(\mathbb{R})$ is defined by the system of norms
$$\Vert \phi \Vert^{2}_{r} = \langle \phi \vert(1 + a^{*}a)^{r}\vert
\phi \rangle $$ for any $r \in  \mathbb{N}$ (see Theorem V.13 of
[\cite{RS72}]).\newline In our case of a rigged Hilbert space
$\Phi \subset  {\mathcal H} \subset \Phi ^{\prime }$, $\Phi $ is
the set of vectors of the form
$$ \phi = \sum_{n=0}^{\infty} c_{n}\vert n\rangle $$
satisfying
$$\Vert \phi \Vert^{2}_{r} = \langle \phi \vert(1 + a^{*}a)^{r}\vert
\phi \rangle = \sum_{n=0}^{\infty}(1 + n^{2})^{r}\vert
c_{n}\vert^{2}
 < \infty $$
for any $r \in \mathbb{N}$.  $\Phi$ is a countably normed space.
Since
$$ \sum_{n=0}^{\infty}(1 + n^{2})^{r}\vert \lambda \vert^{n}/n! < \infty,$$
$e^{\lambda \bar{u}} \in \Phi$ for any $\lambda \in \mathbb{C}$, the
complex numbers.\newline For $|q| < 1$ we have
$$\Vert \sum_{n=0}^{\infty} q^{n}\vert n\rangle_{0} \otimes
\vert n\rangle_{1}\Vert^{2}=\sum_{n=0}^{\infty} q^{2n} =
 \frac{1}{1 - q^{2}},$$
but for $q=1$ this series is divergent:
$$\Vert\sum_{n=0}^{\infty} \vert n\rangle_{0} \otimes
\vert n\rangle_{1}\Vert^{2} = \sum_{n=0}^{\infty} 1 = \infty .$$ In
this sense the vector\newline
$$ \pi^{-1/2} e^{\bar{v}_{0}\bar{v}_{1}} = \pi^{-1/2}
\sum_{n=0}^{\infty} \frac{(\bar{v}_{0}\bar{v}_{1})^{n}}{n!} =
\pi^{-1/2} \sum_{n=0}^{\infty} \vert n\rangle_{0} \otimes \vert
n\rangle _{1}$$ is a generalized vector.  The unitary operator
$e^{iH_{{\rm hbs \,}}}$ sends the generalized vector
$$\sum_{n=0}^{\infty} (D(-\alpha)\vert n\rangle_{0})\otimes \vert
n\rangle_{1} = e^{-\vert \alpha \vert^{2}/2}
e^{\bar{v}_{0}\bar{v}_{1}}
 e^{-\bar{\alpha} \bar{v}_{1}} e^{\alpha \bar{v}_{0}}$$
to the generalized vector\newline
$$ = e^{-(x_{-}^{2}+p_{+}^{2})/2} e^{-\bar{u}_{0}^{2}/2}
e^{\bar{u}_{1}^{2}/2} e^{\sqrt{2}x_{-}\bar{u}_{0}}
e^{i\sqrt{2}p_{+}\bar{u}_{1}} = \pi^{1/2} \vert x_{-}\rangle \otimes
\vert p_{+}\rangle ,$$ where $\alpha = x_{-} + ip_{+}$ and
$$ \vert x_{-}\rangle = \pi^{-1/4} e^{-x_{-}^{2}/2} e^{-\bar{u}_{0}^{2}/2}
e^{\sqrt{2}x_{-}\bar{u}_{0}}$$ is the generalized eigen-state of the
operator
$$ x_{0} = \frac{1}{\sqrt{2}} \left(\frac{\partial}{\partial \bar{u}_{0}}
+ \bar{u}_{0}\right) $$ with eigen-value $x_{-}$, and
$$\vert p_{+}\rangle = \pi^{-1/4} e^{-p_{+}^{2}/2} e^{\bar{u}_{1}^{2}/2}
e^{i\sqrt{2}p_{+}\bar{u}_{1}}$$ is the generalized eigen-state of
the operator
$$p_{1} = \frac{1}{\sqrt{2}i} \left(\frac{\partial}{\partial \bar{u}_{1}}
- \bar{u}_{1}\right) $$ with eigen-value $p_{+}$.  In the usual
Schr\"odinger representation, $\vert p_{+}\rangle $ is represented
by $(2\pi )^{-1/2} e^{ip_{+}x}$ and is a generalized eigen-vector of
$p = -id/dx$ as explained above. Since ${\mathcal H} $ is a
separable Hilbert space, ${\mathcal H} $ has a countable basis. But
for the completeness, ${\mathcal H} $ must have uncountably many
orthogonal generalized vectors.
\subsection{Proofs of theorems 1 - 4, Section 4}
\subsubsection{Proof of Theorem 1}
First, consider the transformation
$$ b_{j} = a_{j} + f_{j}, \  b_{j}^{\dagger} = a_{j}^{\dagger} + \bar{f}_{j}$$
for complex numbers $f_{j}$.  We want to have a unitary
operator $U$ such that
$$ b_{j}U = Ua_{j}, \  b_{j}^{\dagger}U = Ua_{j}^{\dagger} \  \Longleftrightarrow  \  b_{j} = Ua_{j}U^{\dagger}, \  b_{j}^{\dagger} = Ua_{j}^{\dagger}U^{\dagger} .$$
We assume that the operator $U$ is defined by a kernel $U(\bar{u},v)$, that is,
$$(Ug)(\bar{u}) = \int U(\bar{u},v) g(\bar{v}) \prod_{j=1}^{n} e^{-\bar{v}_{j}v_{j}} \frac{d\bar{v}  _{j}dv_{j}}{2\pi \I}.$$
Then we have, by integration by part,
$$\int U(\bar{u},v) \left\{\frac{\partial}{\partial \bar{v}_{j}} g(\bar{v} )\right\} \prod_{j=1}^{n} e^{-\bar{v}_{j}v_{j}} \frac{d\bar{v}_{j}dv_{j}}{2\pi \I}=$$
$$
  -\int g(\bar{v} )\frac{\partial}{\partial \bar{v}_{j}} \left\{U(\bar{u} ,v) \prod_{j=1}^{n} e^{-\bar{v}  _{j}v_{j}}\right\} \frac{d\bar{v}_{j}dv_{j}}{2\pi \I}
= \int v_{j}U(\bar{u}  , v) g(\bar{v}  ) \prod _{j=1}^{n} e^{-\bar{v}  _{j}v_{j}} \frac{d\bar{v}  _{j}dv_{j}}{2\pi i},$$
and similarly
$$ \int U(\bar{u},v) \bar{v}_{j}g(\bar{v}) \prod _{j=1}^{n} e^{-\bar{v}  _{j}v_{j}} \frac{d\bar{v}  _{j}dv_{j}}{2\pi i}$$
$$= -\int U(\bar{u}  , v) \left\{  \frac{\partial }{\partial v_{j}} g(\bar{v}  ) \prod _{j=1}^{n} e^{-\bar{v}  _{j}v_{j}} \right\}  \frac{d\bar{v}  _{j}dv_{j}}{2\pi i} $$
$$
 = \int  \left\{  \frac{\partial }{\partial v_{j}} U(\bar{u}  , v) \right\}  g(\bar{v}  ) \prod _{j=1}^{n} e^{-\bar{v}  _{j}v_{j}} \frac{d\bar{v}  _{j}dv_{j}}{2\pi i}.$$
Thus, it is clear that one should have the following
correspondence:
$$ a_{j}U \leftrightarrow \frac{\partial}{\bar{u}_{j}} U(\bar{u}, v), \  a_{j}^{\dagger}U \leftrightarrow \bar{u}  _{j} U(\bar{u}, v),$$
$$ Ua_{j} \leftrightarrow v_{j} U(\bar{u}, v), \  Ua_{j}^{\dagger} \leftrightarrow \frac{\partial }{\partial v_{j}} U(\bar{u} ,v).$$
If we assume that the kernel of the operator $U$ has the
form
$$U(\bar{u}, v) = c \exp \sum_{j=1}^{n}(\bar{u}_{j}v_{j} + v_{j}\phi_{j} + \bar{u}_{j}\psi_{j})$$
for complex numbers $\phi_{j}$ and $\psi _{j}$, then we
get indeed
$$ b_{j}U(\bar{u}, v) = \left(\frac{\partial }{\partial \bar{u}_{j}} + f_{j}\right) U(\bar{u}, v) = v_{j}U(\bar{u}, v),$$
$$ b_{j}^{*}U(\bar{u},v) = (\bar{u}_{j} + \bar{f}_{j})U(\bar{u}, v) = \frac{\partial }{\partial v_{j}} U(\bar{u},v).$$
Since
$$ \frac{\partial}{\partial \bar{u}_{j}} U(\bar{u},v) =(v_{j} + \psi_{j})U(\bar{u}, v),$$
$$\frac{\partial }{\partial v_{j}} U(\bar{u}, v) = (\bar{u}_{j} + \phi_{j})U(\bar{u},v)$$
equating the coefficients for $\bar{u}_{j}$ and $v_{j}$
we find $\psi_{j} = - f_{j}$ and $\phi _{j} = \bar{f}_{j}$. Thus the result is
$$U(\bar{u},v) = c \exp \sum_{j=1}^{n} (\bar{u}_{j}v_{j} + v_{j}\bar{f}_{j} - \bar{u}_{j}f_{j}).$$
Applying $U$ to the vacuum $f(\bar{v}) = 1$ produces
$$F(\bar{u}) = U\vert 0\rangle= \int  c \exp \sum _{j=1}^{n} (\bar{u}  _{j}v_{j} + v_{j}\bar{f}  _{j} - \bar{u}  _{j}f_{j})  \prod _{j=1}^{n} e^{-\bar{v}  _{j}v_{j}} \frac{d\bar{v}  _{j}dv_{j}}{2\pi i}$$
$$       = c \exp \left\{ - \sum _{j=1}^{n} \bar{u}  _{j}f_{j}\right\}  \int  \exp \left\{  \sum _{j=1}^{n} (\bar{u}  _{j} + \bar{f}  _{j})v_{j}\right\}  \prod _{j=1}^{n} e^{-\bar{v}  _{j}v_{j}} \frac{d\bar{v}  _{j}dv_{j}}{2\pi i}$$
$$       = c \exp \left\{ - \sum _{j=1}^{n} \bar{u}  _{j}f_{j}\right\}  .$$
$$       \Vert F\Vert ^{2} = \vert c\vert ^{2} \int \exp \left\{ - \sum _{j=1}^{n} [u_{j}\bar{f}  _{j} + \bar{u}  _{j}f_{j}] \right\}  \prod _{j=1}^{n} e^{-\bar{u}  _{j}u_{j}} \frac{d\bar{u}  _{j}du_{j}}{2\pi i}$$
$$       = \vert c\vert ^{2} \int \exp - \left\{  \sum _{j=1}^{n} [(\bar{u}  _{j} + \bar{f}  _{j})(u_{j} + f_{j}) - \bar{f}  _{j}f_{j}]\right\}  \frac{d\bar{u}  _{j}du_{j}}{2\pi i}$$
$$       = \vert c\vert ^{2} \exp \left\{  \sum _{n=1}^{n}\bar{f}  _{j}f_{j}\right\}  .$$
The normalization of $F$, i.e., $\Vert F\Vert  = 1$,
requires
$$       c = \theta  \exp \left\{ (-1/2)\sum _{j=1}^{n} \bar{f}  _{j}f_{j}\right\} , \  \vert \theta \vert  = 1,$$
and we have the following theorem.\\[2mm]\newline
\subsubsection{Proof of Theorem 2}
Now we consider more general linear canonical
transformation (Bogoliubov transformation\index{Bogoliubov
transformation})
$$       b_{j} = \sum _{k=1}^{n}(\Phi _{jk}a_{k} + \Psi _{jk}a^{*}_{k}), \  b_{j}^{*} = \sum _{k=1}^{n}(\bar{\Phi }  _{jk}a_{k}^{*} + \bar{\Psi }  _{jk}a_{k})$$
satisfying
$$       [b_{j}, b_{k}] = [b_{j}, b_{k}] = 0, \  [b_{j}, b_{k}^{*}] = \delta _{jk},$$
and find the kernel $U(\bar{u}  , v)$ of the unitary
operator $U$ which implements the linear canonical
transformation
$$       b_{j}U = Ua_{j}, \  b_{j}^{*}U = Ua_{j}^{*}.$$
From the above commutation relations, we have
$$      0 = [b_{j}, b_{k}] = \sum _{i=1,m=1}^{n}\left[  \Phi _{jl}a_{l} + \Psi _{jl}a^{*}_{l}, \Phi _{km}a_{m} + \Psi _{km}a^{*}_{m}\right] $$
$$      = \sum _{l=1, m=1}^{n}\left(  \Phi _{jl}\Psi _{km}[a_{l}, a^{*}_{m}] + \Psi _{jl}\Phi _{km}[a^{*}_{l}, a_{m}]\right) $$
$$      = \sum _{l=1, m=1}^{n}\left(  \Phi _{jl}\Psi _{km}\delta _{lm} - \Psi _{jl}\Phi _{km}\delta _{lm}\right)  = \sum _{m=1}^{n}\left(  \Phi _{jm}\Psi _{km} - \Psi _{jm}\Phi _{km}\right) $$
$$      \delta _{jk} = [b_{j}, b_{k}^{*}] = \sum _{i=1,m=1}^{n}\left[  \Phi _{jl}a_{l} + \Psi _{jl}a^{*}_{l}, \bar{\Phi } _{km}a_{m}^{*} + \bar{\Psi } _{km}a_{m}\right] $$
$$      = \sum _{i=1,m=1}^{n}\left(  \Phi _{jl}\bar{\Phi } _{km}[a_{l}, a_{m}^{*}] + \Psi _{jl}\bar{\Psi } _{km}[a^{*}_{l}, a_{m}]\right) $$
$$      = \sum _{i=1,m=1}^{n}\left(  \Phi _{jl}\bar{\Phi } _{km}\delta _{lm} - \Psi _{jl}\bar{\Psi } _{km}\delta _{lm}\right)  = \sum _{m=1}^{n}\left(  \Phi _{jm}\bar{\Phi } _{km} - \Psi _{jm}\bar{\Psi } _{km}\right) .$$
These relations can be described by the matrix notation:
\begin{equation} \label{4.3}
   0 = \Phi \Psi ^{T} - \Psi \Phi ^{T}, \  I = \Phi \Phi ^{*} -
\Psi \Psi ^{*}.
\end{equation}
We want to have a unitary operator $U$ such that
$$      b_{j}U = Ua_{j}, \  b_{j}^{*}U = Ua_{j}^{*}.$$
We assume that the kernel of the operator $U$ has the form
$$      U(\bar{u} , v) = c \exp \left(  \frac{1}{2} (v \  \bar{u} ) \left( \begin{array}{ccc} A^{11}&A^{12}\\{} A^{21}&A^{22}\end{array} \right)
\left( \begin{array}{c} v\\{} \bar{u} \end{array} \right)
\right) , \  A^{jk} = {}^tA^{kj}.$$ Then we have
$$      b_{j}U(\bar{u} , v) = \sum _{k=1}^{n}(\Phi _{jk}\frac{\partial }{\partial \bar{u} _{k}} + \Psi _{jk}\bar{u} _{k})U(\bar{u} , v) = v_{j}U(\bar{u} , v) .$$
Since
$$      \frac{\partial }{\partial \bar{u} _{k}} U(\bar{u} , v) =  \sum _{m=1}^{n}(A^{21}_{km}v_{m} + A^{22}_{km}\bar{u} ) U(\bar{u} , v),$$
equating the coefficients for $\bar{u} _{l}$ and $v_{l}$ we
have
$$      \Phi A^{22} + \Psi  = 0, \  \Phi A^{21} = I.$$
In the same way, from the equation
$$      b_{j}^{*}U(\bar{u} , v) = \sum _{k=1}^{n}(\bar{\Phi } _{jk}a_{k}^{*} + \bar{\Psi } _{jk}a_{k}) U(\bar{u} , v) = \frac{\partial }{\partial v_{j}} U(\bar{u} , v),$$
we deduce
$$      \bar{\Psi } A^{22} + \bar{\Phi }  = A^{12}, \  \bar{\Psi } A^{21} = A^{11}.$$
This gives
$$      A^{22} = -\Phi ^{-1}\Psi , \  A^{21} = \Phi ^{-1}, \  A^{11} = \bar{\Psi } \Phi ^{-1}.$$
Applying $U$ to the vacuum $\vert 0\rangle $, we have
$$      F(\bar{u} ) = U\vert 0\rangle  = c \int \exp \left(  \frac{1}{2} (v \  \bar{u} ) \left( \begin{array}{ccc} A^{11}&A^{12}\\{} A^{21}&A^{22}\end{array} \right) \left( \begin{array}{c} v\\{} \bar{u} \end{array} \right) \right)  \prod _{j=1}^{n} e^{-\bar{v} _{j}v_{j}} \frac{d\bar{v} _{j}dv_{j}}{2\pi i}$$
$$      = c \exp (1/2)\{ -\bar{u} \Phi ^{-1}\Psi \bar{u} \}  \int \exp (1/2)\{ 2\bar{u} \Phi ^{-1}v + v\bar{\Psi } \Phi ^{-1}v -\bar{v} v\}  \prod _{j=1}^{n}\frac{d\bar{v} _{j}dv_{j}}{2\pi i}.$$
The integral is
$$      \int \exp (-1/2)\left\{ (v \  \bar{v} )\left( \begin{array}{ccc} -\bar{\Psi } \Phi ^{-1}&I\\{} I&0\end{array} \right) \left( \begin{array}{c} v\\{} \bar{v} \end{array} \right) + 2\bar{u} \Phi ^{-1}v\right\}  \prod _{j=1}^{n}\frac{d\bar{v} _{j}dv_{j}}{2\pi i}$$
$$      = \left[ \det \left( \begin{array}{ccc} I&\bar{\Psi } \Phi ^{-1}\\{} 0&I\end{array} \right) \right] ^{-1/2}$$
$$      \times  \exp \left[ (1/2)(0 \  2\bar{u} \Phi ^{-1})\left( \begin{array}{ccc} 0&I\\{} I&\bar{\Psi } \Phi ^{-1}\end{array} \right) \left( \begin{array}{c} 0\\{} 2\bar{u} \Phi ^{-1}\end{array} \right) \right]  = 1,$$
where we used the following formula of Gaussian integral:
$$      \int \exp \left(  \frac{-1}{2} (v \  \bar{v} ) \left( \begin{array}{ccc} A^{11}&A^{12}\\{} A^{21}&A^{22}\end{array} \right) \left( \begin{array}{c} v\\{} \bar{v} \end{array} \right)  + (f_{1} \  f_{2})\left( \begin{array}{c} v\\{} \bar{v} \end{array} \right) \right)  \prod _{j=1}^{n}\frac{d\bar{v} _{j}dv_{j}}{2\pi i}$$
$$      = \left[ \det \left( \begin{array}{ccccc} A^{21}&A^{22}\\{} A^{11}&A^{12}\end{array} \right) \right] ^{-1/2} \exp (1/2)\left\{ (f_{1} \  f_{2})\left( \begin{array}{ccc} A^{11}&A^{12}\\{} A^{21}&A^{22}\end{array} \right) ^{-1}\left( \begin{array}{c} f_{1}\\{} f_{2}\end{array} \right) \right\} .$$
Now we calculate
$$      \Vert F\Vert ^{2} = \vert c\vert ^{2} \int \exp -\{ (1/2)(u\bar{\Phi } ^{-1}\bar{\Psi } u + \bar{u} \Phi ^{-1}\Psi \bar{u} ) + \bar{u} u\}  \prod _{j=1}^{n}\frac{d\bar{u} _{j}du_{j}}{2\pi i}$$
$$      = \vert c\vert ^{2} \int \exp \left\{  \frac{-1}{2} (u \  \bar{u} ) \left( \begin{array}{ccc} \bar{\Phi } ^{-1}\bar{\Psi } &I\\{} I&\Phi ^{-1}\Psi \end{array} \right) \left( \begin{array}{c} u\\{} \bar{u} \end{array} \right) \right\}  \prod _{j=1}^{n}\frac{d\bar{u} _{j}du_{j}}{2\pi i}$$
$$      = \vert c\vert ^{2} \left[ \det \left( \begin{array}{ccc} I&\Phi ^{-1}\Psi \\{} \bar{\Phi } ^{-1}\bar{\Psi } &I\end{array} \right) \right] ^{-1/2}.$$
$$      \det \left( \begin{array}{ccccccc} I&\Phi ^{-1}\Psi \\{} \bar{\Phi } ^{-1}\bar{\Psi } &I\end{array} \right)  = \det \left[ \left( \begin{array}{ccccc} I&0\\{} -\bar{\Phi } ^{-1}\bar{\Psi } &I\end{array} \right) \left( \begin{array}{ccc} I&\Phi ^{-1}\Psi \\{} \bar{\Phi } ^{-1}\bar{\Psi } &I\end{array} \right) \right] $$
$$      = \det \left[ \left( \begin{array}{ccccccc} I&-\Phi ^{-1}\Psi \\{} 0&I\end{array} \right) \left( \begin{array}{ccccc} I&\Phi ^{-1}\Psi \\{} \bar{\Phi } ^{-1}\bar{\Psi } &I\end{array} \right) \right]  = \det \left( \begin{array}{ccc} I&-\Phi ^{-1}\Psi \bar{\Phi } ^{-1}\bar{\Psi } \\{} 0&I\end{array} \right)  $$
$$      = \det (I -\Phi ^{-1}\Psi \bar{\Phi } ^{-1}\bar{\Psi } ) = \det (I - \Phi ^{-1}\Psi \Psi ^{*}\Phi ^{*-1})$$
$$      = \det (I - \Phi ^{-1}(\Phi \Phi ^{*} - I)\Phi ^{*-1}) = \det (\Phi ^{*}\Phi )^{-1},$$
where we have the relation (\ref{4.3}).  The constant $c$
is calculated to be
$$      c = \theta  (\det \Phi \Phi ^{*})^{-1/4}, \  \vert \theta \vert  = 1,$$
and we get the following result.\newline
\subsubsection{Proof of Theorem 3}
Proof.  Consider the operator
\begin{equation} \label{4.5}
  a_{j}(t) = e^{itH}a_{j}e^{-itH}, \  a_{j}^{*}(t) =
e^{itH}a_{j}^{*}e^{-itH}.
\end{equation}
Differentiating with respect to $t$, we have
$$      \frac{1}{i}\frac{d a_{j}(t)}{dt} = [H, a_{j}(t)] = [H(t), a_{j}(t)] = e^{itH}[H, a_{j}]e^{-itH}$$
$$      = e^{itH} \{ -f_{j}\} e^{-itH} = - f_{j},$$
$$      \frac{1}{i}\frac{d a_{j}^{*}(t)}{dt} = [H, a_{j}^{*}(t)] = [H(t), a_{j}^{*}(t)] = e^{itH}[H, a_{j}^{*}]e^{-itH}$$
$$      = e^{itH} \bar{f} _{j} e^{-itH} = \bar{f} _{j}.$$
Integrating this system, we have
$$      a_{j}(t) = a_{j} - itf_{j}, \  a_{j}^{*}(t) = a_{j}^{*} + it\bar{f} _{j}.$$
Thus (\ref{4.5}) is a linear canonical transformation of
Theorem \ref{thm4.1}. Therefore the kernel $U(\bar{u} , v)$
has the form (\ref{4.4}). In order to find $c$ precisely,
we differentiate $U(\bar{u} , v)$ and $e^{itH}$ and compare
them.
$$      \frac{1}{i} \frac{d}{dt} U(\bar{u} , v) = \left(  \frac{1}{i} \frac{dc}{dt} + c \sum _{j=1}^{n} (f_{j}\bar{u} _{j} + \bar{f} _{j}v_{j})\right)  U(\bar{u} , v)$$
$$      = c \sum _{j=1}^{n} \left( f_{j} \bar{u} _{j} + \bar{f} _{j} \frac{\partial }{\partial \bar{u} _{j}}\right) U(\bar{u} , v) = c ( \sum _{j=1}^{n} (f_{j}\bar{u} _{j} + \bar{f} _{j}v_{j}) +it \bar{f} _{j}f_{j}) U(\bar{u} , v)$$
As a result we obtain
$$      \frac{1}{i} \frac{dc}{dt} = itc ( \sum _{j=1}^{n} \bar{f} _{j}f_{j}), \  {\rm and \, }\
c = \exp \left\{ - \frac{1}{2} t^{2} \sum _{j=1}^{n}
\bar{f} _{j}f_{j}\right\} .$$ \hfill $ $

\subsubsection{Proof of Theorem 4}
Proof. Consider the operator\newline \indent
\begin{equation} \label{4.9}
  a_{j}(t) = e^{itH}a_{j}e^{-itH}, \  a_{j}^{*}(t) =
e^{itH}a_{j}^{*}e^{-itH}.
\end{equation}
Differentiating with respect to $t$, we have
$$      \frac{1}{i}\frac{d a_{j}(t)}{dt} = [H, a_{j}(t)] = [H(t), a_{j}(t)] = e^{itH}[H, a_{j}]e^{-itH}$$
$$      = e^{itH} \sum _{k=1}^{n} - \{ C_{jk}a_{k} + B_{jk}a_{k}^{*}\} e^{-itH} = - \sum _{k=1}^{n} \{ C_{jk}a_{k}(t) + B_{jk}a_{k}^{*}(t)\} ,$$
$$      \frac{1}{i}\frac{d a_{j}^{*}(t)}{dt} = [H, a_{j}^{*}(t)] = [H(t), a_{j}^{*}(t)] = e^{itH}[H, a_{j}^{*}]e^{-itH}$$
$$      = e^{itH} \sum _{k=1}^{n} \{ \bar{B} _{jk}a_{k} + \bar{C} _{jk}a_{k}^{*}\} e^{-itH} = \sum _{k=1}^{n} \{ \bar{B} _{jk}a_{k}(t) + \bar{C} _{jk}a_{k}^{*}(t)\} .$$
Integrating this system, we have
$$      a_{j}(t) = \sum _{k=1}^{n}(\Phi _{jk}a_{k} + \Psi _{jk}a^{*}_{k}), \  a_{j}^{*}(t) = \sum _{k=1}^{n}(\bar{\Phi } _{jk}a_{k}^{*} + \bar{\Psi } _{jk}a_{k}),$$
where
$$      \left( \begin{array}{ccccc} \Phi &\Psi \\{} \bar{\Psi } &\bar{\Phi } \end{array} \right)  = \exp \left\{ it\left( \begin{array}{ccc} -C&-B\\{} \bar{B} &\bar{C} \end{array} \right) \right\} .$$
Thus, Eq. (\ref{4.9}) is a linear canonical transformation. In
order to find $c$ precisely, we differentiate $U(\bar{u} ,
v)$ and $e^{itH}$ and compare the results.
$$      \frac{1}{i} \frac{d}{dt} U(\bar{u} , v) = \left(  \frac{1}{i} \frac{dc}{dt} + c \frac{1}{2i} (v \bar{u} )\frac{d}{dt} \left( \begin{array}{ccc} A^{11}&A^{12}\\{} A^{21}&A^{22}\end{array} \right) \left( \begin{array}{c} v\\{} \bar{u} \end{array} \right)  \right) U(\bar{u} , v)$$
$$      = c \frac{1}{2} \left( \bar{u} B\bar{u}  + \frac{\partial }{\partial \bar{u} }\bar{B} \frac{\partial }{\partial \bar{u} } + 2\bar{u} C\frac{\partial }{\partial \bar{u} }\right) U(\bar{u} , v).$$
Put $v = \bar{u}  = 0$.  Then we have
$$      \frac{1}{i} \frac{dc}{dt} = c \frac{1}{2} \sum _{ij}\bar{B} _{ij}A^{22}_{ij} = c \frac{1}{2} \sum _{ij}A^{22}_{ji}\bar{B} _{ij} = c \frac{1}{2} {\rm Tr\, }A^{22}B = - c \frac{1}{2} {\rm Tr\, }(\Phi ^{-1}\Psi B).$$
We shall seek $c$ in the form $c = (\det M)^{-1/2}$.  From the
formula $\det M = e^{{\rm Tr\, }\log M}$, we find \begin{multline*}
 \frac{d}{dt} \det M = \left(  \frac{d}{dt} {\rm Tr\, }\log
M\right) \det M = {\rm Tr\, }\left( \frac{d}{dt} \log M \right) \det
M =\\ {\rm Tr\, }\left( M^{-1} \frac{dM}{dt}\right) \det
M.\end{multline*} Thus we obtain the equation for $M$:
$$      \frac{1}{i} {\rm Tr\, }\left( M^{-1} \frac{dM}{dt}\right)  = {\rm Tr\, }(\Phi ^{-1}\Psi B).$$
Recalling that the operators $\Phi , \Psi $ are the
solution of the equation
$$      \frac{1}{i} \frac{d}{dt} \left( \begin{array}{ccccccc} \Phi &\Psi \\{} \bar{\Psi } &\bar{\Phi } \end{array} \right)  = \left( \begin{array}{ccccc} \Phi &\Psi \\{} \bar{\Psi } &\bar{\Phi } \end{array} \right) \left( \begin{array}{ccc} -C&-B\\{} \bar{B} &\bar{C} \end{array} \right) ,$$
we have
$$      \frac{1}{i} \Phi ^{-1} \frac{d\Phi }{dt} = -C + \Phi ^{-1}\Psi \bar{B} .$$
We now set $M = \Phi e^{itC}$, then we have
$$      \frac{1}{i} {\rm Tr\, }\left( M^{-1} \frac{dM}{dt}\right)  = {\rm Tr\, }\left( e^{-itC} \Phi ^{-1} \left( \frac{1}{i}\frac{d\Phi }{dt} e^{itC} + \Phi Ce^{itC}\right) \right) $$
$$      = {\rm Tr\, }\left( e^{-itC} \Phi ^{-1} \frac{1}{i}\frac{d\Phi }{dt} e^{itC} + e^{-itC}Ce^{itC}\right) $$
$$      = {\rm Tr\, }\left( e^{-itC} (-C + \Phi ^{-1}\Psi \bar{B} ) e^{itC} + e^{-itC}Ce^{itC}\right)  = {\rm Tr\, }(\Phi ^{-1}\Psi \bar{B} ).$$
Thus $c = (\det \Phi e^{itC})^{-1/2}$.  Therefore the
kernel $U(\bar{u} , v)$ has the form (\ref{4.7}),
(\ref{4.8}). \hfill $ $ \vskip 12pt \noindent
\subsection{Proofs for Section 6}
\subsubsection{Generalized Bell basis} First we show completeness:
$$\pi^{-1} \int e^{-\bar{\alpha}\alpha} e^{\bar{u}_{0}\bar{u}_{1}}e^{- \bar{\alpha} \bar{u}_{1}} e^{\alpha \bar{u}_{0}} e^{v_{0}v_{1}}e^{- \alpha v_{1}} e^{\bar{\alpha} v_{0}} dx_{-} dp_{+}$$
$$= \pi^{-1} \int e^{-\bar{\alpha} \alpha} e^{\bar{u}_{0}\bar{u}_{1}} e^{\bar{\alpha}(v_{0} - \bar{u}_{1})} e^{\alpha(\bar{u}_{0}- v_{1})}e^{v_{0} v_{1}}dx_{-}dp_{+}$$
$$= \int e^{-\bar{\alpha} \alpha} e^{\bar{u}_{0}\bar{u}_{1}} e^{\bar{\alpha}(v_{0}-\bar{u}_{1})}e^{\alpha (\bar{u}_{0}-v_{1})}e^{v_{0} v_{1}}\frac{d\bar{\alpha} d\alpha}{2\pi \I}$$
$$= e^{\bar{u}_{0}\bar{u}_{1}} e^{(\bar{u}_{0}-v_{1})(v_{0}-\bar{u}_{1})}e^{v_{0} v_{1}}=e^{\bar{u}  _{0}v_{0}} e^{\bar{u}_{1}v_{1}}.$$
This is the kernel of the identity operator (see (\ref{4.1})). Thus (\ref{5.1}) is a complete system.

In order to show that (\ref{5.1}) is an orthogonal system, we
rewrite the vector $\sum_{n=0}^{\infty}(D(\alpha)\vert n\rangle_{0})\otimes \vert n\rangle_{1}$ of (\ref{5.1}) as
$$ e^{-\vert \alpha \vert^{2}/2} e^{\bar{v}_{0}\bar{v}_{1}} e^{-\bar{\alpha} \bar{v}_{1}}
e^{\alpha \bar{v}_{0}} = e^{-\vert \alpha \vert^{2}/2} e^{\bar{v}
_{0}\bar{v}_{1}} e^{- (x_{-} - \I p_{+}) \bar{v}_{1}}
e^{(x_{-}+\I p_{+})\bar{v}_{0}}$$
$$= e^{-\vert \alpha \vert^{2}/2} e^{\bar{v}_{0}\bar{v}_{1}} e^{x_{-}(\bar{v}_{0} - \bar{v}_{1})}
e^{\I p_{+}(\bar{v}_{0} + \bar{v}_{1})} =$$ $$= e^{-\vert \alpha \vert
^{2}/2} e^{-(\bar{v}_{0}-\bar{v}_{1})^{2}/4}
e^{(\bar{v}_{0}+\bar{v}_{1})^{2}/4} e^{x_{-}(\bar{v}_{0} - \bar{v}
_{1})} e^{\I p_{+}(\bar{v}_{0} + \bar{v}_{1})}.$$

\subsubsection{Proof of Orthogonality relation (\ref{orth-rel})} This is a straightforward calculation:
$$ \langle x_{-}^{\prime}\vert x_{-}\rangle=\pi^{-1/2} e^{-x_{-}^{\prime 2}/2} e^{-x_{-}^{2}/2} \int e^{-u_{0}^{2}/2} e^{\sqrt{2}x_{-}^{\prime}u_{0}} e^{-\bar{u}_{0}^{2}/2} e^{\sqrt{2}x_{-}\bar{u}_{0}} e^{-\bar{u}_{0}u_{0}}\frac{d\bar{u}_{0} du_{0}}{2\pi \I}$$
$$= \pi^{-1/2} e^{-x_{-}^{\prime 2}/2} e^{-x_{-}^{2}/2} \int e^{-2q^{2}} e^{\sqrt{2}q(x_{-}^{\prime} + x_{-})} e^{\I\sqrt{2}p(x_{-}^{\prime} - x_{-})} \frac{dq dp}{\pi}$$
$$=\pi ^{-1/2} e^{-x_{-}^{\prime 2}/2} e^{-x_{-}^{2}/2} \int e^{-(q^{2}-q(x_{-}^{\prime } + x_{-}))} e^{\I p(x_{-}^{\prime}-x_{-})} \frac{dq dp}{2\pi }$$
$$= \pi^{-1/2} e^{-x_{-}^{\prime 2}/2} e^{-x_{-}^{2}/2} \sqrt{\pi} e^{(x_{-}^{\prime} + x_{-})^{2}/4} \delta (x_{-}^{\prime}-x_{-})=\delta(x_{-}^{\prime}-x_{-}),$$
and in a similar way
$$\langle p_{+}^{\prime }\vert p_{+}\rangle=\delta (p_{+}^{\prime}-p_{+}).$$
\end{appendix}


\begin{thebibliography}{99}

\bibitem{Ba61}
V. Bargmann:
\newblock On a {H}ilbert {S}pace of {A}nalytic {F}unctions
and an {A}ssociated {I}ntegral {T}ransform, {P}art {I}.
 Commun. Pure Appl. Math. \textbf{14}, 187--214 (1961)
%
\bibitem{Be93}
C.H. Bennett,  G.~Brassard, C. Cr{\'e}peau, R. Jozsa,
A.~Peres and W.K. Wootters:
\newblock Teleporting unknown quantum state via dual
classical and   {E}instein-{P}odolsky-{R}osen channels.
 Phys. Rev. Lett. \textbf{70}, 1895 (1993)
%
\bibitem{Be66}
F.A. Berezin:
\newblock {\em The {M}ethod of {S}econd {Q}uantization}.
\newblock (Academic Press, New York 1966)
%
\bibitem{Bo97}
D. Bouwmeester,  J.-W. Pan, K. Mattle, M. Eibl,
H.~Weinfurter and A.~Zeilinger:
\newblock Experimental quantum teleportation.
 Nature (London) \textbf{390}, 575 (1997)
%
\bibitem{Bo03}
W.P. Bowen, N.~Treps, B.C. Buchler, R. Schnabel, C.R.
Timothy, H.-A. Bachor, T. Symul, P.K.~Lam and R.~Timothy:
\newblock Experimental investigation of continuous-variable
quantum teleportation.
\newblock  Phys. Rev. A \textbf{67}, 032302 (2003)
%
\bibitem{BK98}
S.L. Braunstein and H.J. Kimble:
\newblock Teleportation of {C}ontinuous {Q}uantum
{V}ariables.
\newblock Phys. Rev. Lett. \textbf{80}, 869--872 (1998)
%
\bibitem{GZ04}
 C.W. Gardiner and P.~Zoller:
\newblock {\em Quantum {N}oise}.
\newblock (Springer 2004)
%
\bibitem{FS80}
L.D. Faddeev and A.A. Slanov:
\newblock {\em Gauge {F}ields: {I}ntroduction to {Q}uantum
{T}heory}.
\newblock (Benjamin/Cummings Publishing Co. Inc.,
London-Amsterdam-Sydney-Tokyo 1980)
%
\bibitem{Fu03}
M. Fujii:
\newblock Continuous-{V}ariable {Q}uantum {T}eleportation
with a {C}onventional{L}aser.
\newblock {\em quant-ph/0304148}, (2003)
%
\bibitem{Fu98}
A. Furusawa, J.L. S{\o}rensen, S.L. Braunstein, C.A. Fuchs,
H.J.~Kimble and E.S. Polzik:
\newblock Unconditional {Q}uantum {T}eleportation.
\newblock {Science} \textbf{282}, 706--709 (1998)
%
\bibitem{PM06}
S. Pirandola and S.~Mancini:
\newblock Quantum {T}eleportation with {C}ontinuous
{V}ariables: a survey.
\newblock {\em quant-ph/0604027 v2} (2006)
%
\bibitem{RS01}
T. Rudolph and B.C. Sanders:
\newblock Requrement of {O}ptical {C}oherence for
{C}ontinuous-{V}ariable {Q}uantum {T}eleportation.
\newblock {Phys. Rev. Lett.} \textbf{87}, 077903--1 -- 077903--4
(2001)
%
\bibitem{SZ97}
M.O. Scully and M.S. Zubairy:
\newblock {\em Quantum {O}ptics}.
\newblock (Cambridge University Press, Cambridge 1997)
%
\bibitem{Se62}
I.E. Segal:
\newblock Mathematical characterization of the physical
vacuum for a linear
  {B}ose-{E}instein field. (Foundations of the Dynamics of
  Infinite Systems. III).
  \newblock {Illinois J. Math.} \textbf{6}, 500--525 (1962)
%
\bibitem{Se78}
I.E. Segal:
\newblock The complex-wave representation of the free boson
field.
\newblock {Adv. in Math. Suppl. Stud.} \textbf{3}, 500--525
(1978)
%
\bibitem{Ta05}
N. Takei, H.~Yonezawa, T.~Aoki and A.~Furusawa:
\newblock High-fidelity teleportation beyond the no-cloning
limit and entanglement for continuous variables.
\newblock {Phys. Rev. Lett.} \textbf{96}, 220502--1--4
(2005)
%
\bibitem{Va94}
L. Vaidman:
\newblock Teleportation of quantum states.
\newblock {Phys. Rev. A} \textbf{49}, 1473 -- 1476 (1994)
%
\bibitem{EF02}
S.J. van Enk and C.A. Fuchs:
\newblock Quantum {S}tate of an {I}deal {P}ropagating
{L}aser {F}ield.
\newblock {Phys. Rev. Lett.} \textbf{88}, 027902--1 -- 027902--4
 (2002)
%
\bibitem{GV64}
I.~M Gel'fand and N.~Ya. Vilenkin.
\newblock {\em Applications of {H}armonic {A}nalysis}, volume~4 of {\em
  Generalized Functions}.
\newblock Academic Press, New York and London, 1964.
%
\bibitem{NC00}
M.A. Nielsen and I.L. Chuang.
\newblock {\em Quantum {C}omputation and {Q}uantum {I}nformation}.
\newblock Cambridge {U}niversity {P}ress, 2000.
%
\bibitem{RS72}
M.~Reed and B.~Simon.
\newblock {\em Fourier Analysis, Self-adjointness}, volume~I of {\em Methods of
  Modern Mathematical Physics}.
\newblock Academic Press, New York, San Francisco, London, 1972.

\end{thebibliography}
\end{document}